\documentstyle[12pt,aaspp4,psfig]{article}
\def\m{$\mu$m}
\def\IRAS{{\it IRAS}}
\def\ISO{{\it ISO}}
\def\ISOcolor{${I_\nu (6.75 \mu {\rm m})} \over {I_\nu (15 \mu {\rm m})}$}
\def\ISOcolorb{${I_\nu (15 \mu {\rm m})} \over {I_\nu (6.75 \mu {\rm m})}$}
\def\ISOcolorc{${I_\nu (4.5 \mu {\rm m})} \over {I_\nu (6.75 \mu {\rm m})}$}
\def\ISOIRASa{${I_\nu (6.75 \mu {\rm m})} \over {I_\nu (12 \mu {\rm m})}$}
\def\ISOIRASb{${I_\nu (15 \mu {\rm m})} \over {I_\nu (12 \mu {\rm m})}$}
\def\IRAScolor{${I_\nu (60 \mu {\rm m})} \over {I_\nu (100 \mu {\rm m})}$}

\def\be{\begin{equation}}
\def\ee{\end{equation}}
\def\about{$\sim$}
\def\HI{\ion{H}{1}}
\def\HII{\ion{H}{2}}

\def\NeII{[\ion{Ne}{2}]}
\def\NeIII{[\ion{Ne}{3}]}
\def\Rmira{$R(6.75 \mu{\rm m})$}
\def\Rmirb{$R(15 \mu{\rm m})$}
\def\RB{$R(0.44 \mu{\rm m})_{25}$}
\begin{document}
\title{Mid-Infrared Observations of Normal Star-Forming Galaxies: The Infrared Space Observatory Key Project Sample\footnote{\scriptsize This paper is based on observations with the Infrared Space Observatory (\ISO). \ISO\ is an ESA project with instruments funded by ESA member states (especially the PI countries: France, Germany, The Netherlands and the United Kingdon) and with the participation of ISAS and NASA.}}

\author{Daniel A. Dale,\footnote{IPAC, California Institute of Technology,
MS 100-22, Pasadena, CA 91125}
\addtocounter{footnote}{-1}
 Nancy A. Silbermann,\footnotemark
\addtocounter{footnote}{-1}
 George Helou,\footnotemark
\addtocounter{footnote}{-1}
 Emmanuel Valjavec,\footnotemark
\addtocounter{footnote}{-1}
 Sangeeta Malhotra,\footnotemark
\addtocounter{footnote}{-1}
 Charles A. Beichman,\footnotemark
\addtocounter{footnote}{-1}
 James Brauher,\footnotemark
\addtocounter{footnote}{-1}
 Alessandra Contursi,\footnotemark
 Harriet L. Dinerstein,\footnote{University of Texas, Astronomy Dept. RLM 15.308, Austin, TX 78712}
 David J. Hollenbach,\footnote{NASA/Ames Research Center, MS 245-6, Moffett Field, CA 94035}
 Deidre A. Hunter,\footnote{Lowell Obs., 1400 Mars Hill Rd., Flagstaff, AZ 86001}
\addtocounter{footnote}{-4}
 Sonali Kolhatkar,\footnotemark
\addtocounter{footnote}{3}
 Kwok-Yung Lo,\footnote{University of Illinois, Astronomy Dept. 1002 W. Green St. Urbana, IL 61801}
\addtocounter{footnote}{-5}
 Steven D. Lord,\footnotemark
\addtocounter{footnote}{-1}
 Nanyao Y. Lu,\footnotemark
\addtocounter{footnote}{1}
 Robert H. Rubin,\footnotemark
\addtocounter{footnote}{2}
 Gordon J. Stacey,\footnote{Cornell University, Astronomy Dept., 220 Space Science Building,
 Ithaca, NY 14853}
 Harley A. Thronson, Jr.,\footnote{NASA HQ Code SR, 300 E. St. SW, Washington D.C. 20024}
 Michael W. Werner,\footnote{Jet Propulsion Laboratory, MS 233-303, 4800 Oak Grove Rd.,
Pasadena, CA 91109}
 Harold G. Corwin, Jr.$^{2}$
}

\begin{abstract}
We present mid-infrared maps and preliminary analysis for 61 galaxies observed with the ISOCAM instrument aboard the {\it Infrared Space Observatory}.  Many of the general features of galaxies observed at optical wavelengths---spiral arms, disks, rings, and bright knots of emission---are also seen in the mid-infrared, except the prominent optical bulges are absent at 6.75 and 15 \m.  In addition, the maps are quite similar at 6.75 and 15 \m, except for a few cases where a central starburst leads to lower \ISOcolor\ ratios in the inner region.  We also present infrared flux densities and mid-infrared sizes for these galaxies.  The mid-infrared color \ISOcolor\ shows a distinct trend with the far-infrared color \IRAScolor.  The quiescent galaxies in our sample (\IRAScolor\ $\lesssim 0.6$) show \ISOcolor\ near unity, whereas this ratio drops significantly for galaxies with higher global heating intensity levels.  Azimuthally-averaged surface brightness profiles indicate the extent to which the mid-infrared flux is centrally concentrated, and provide information on the radial dependence of mid-infrared colors.  The galaxies are mostly well resolved in these maps: almost half of them have $<$ 10\% of their flux in the central resolution element.  A comparison of optical and mid-infrared isophotal profiles indicates that the optical flux at 4400 \AA\ near the optical outskirts of the galaxies is approximately eight (seven) times that at 6.75 \m\ (15 \m), comparable with observations of the diffuse quiescent regions of the Milky Way.  
\end{abstract}


\section{Introduction}
Normal galaxies, defined as those with on-going star formation, account for most of the luminous mass in the local Universe.  Their luminosity is derived from stars and they span a broad range of observed morphologies, luminosities, and infrared-to-blue ratios.  Results from \IRAS\ have shown that the infrared colors from the four \IRAS\ bands are sensitive indices of the radiation intensity in the interstellar medium.  The mid-infrared (5-20 \m) emission is dominated by very small grains fluctuating to high temperatures and polycyclic aromatic hydrocarbons (PAHs) (Helou et al. 2000).  These grains are not in thermal equilibrium but they still convert heating photons and thereby trace star formation.  Larger grains in thermal equilibrium dominate the emission from normal galaxies at longer wavelengths. By comparing the mid-infrared emission to other components of the galaxy such as \HI, H$_{2}$, ionized gas, and starlight, one can derive the physical properties of the interstellar gas, dust and radiation field in galaxies (e.g. Vigroux et al. 1999).

The {\it Infrared Space Observatory} (\ISO) U.S. Key Project on Normal Galaxies (PI: G. Helou, proposal id: SF\_GLX\_*, hereafter the ``Key Project'') was proposed to study the interstellar medium of a broad range of normal galaxies using three of the four instruments aboard \ISO: ISOCAM, ISOLWS, and ISOPHOT.  Under NASA guaranteed time, the U.S. Key Project obtained \ISO\ observations of 69 galaxies.  Nine relatively nearby and extended galaxies were chosen to provide spatially resolved cases so that various phases of the interstellar medium could be studied independently.  The remaining 60 galaxies in the Key Project sample cover the full range of observed morphologies, luminosities, and \IRAS\ colors seen in normal galaxies.  Using this diverse sample, we hope to gain new insight into the star formation process on the scale of galaxies, especially its drivers and inhibitors (Helou et al. 1996). 

\section{The U.S. Key Project ISOCAM Sample}
\begin{figure}
\centerline{\psfig{figure=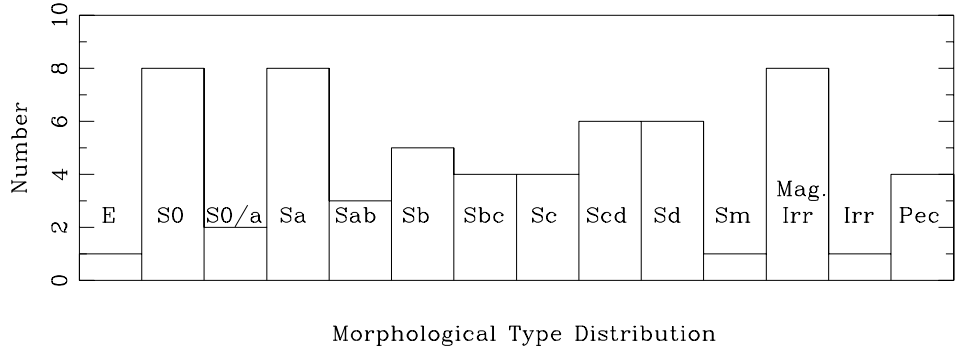,width=5in,bbllx=37pt,bblly=165pt,bburx=565pt,bbury=364pt}}
\caption[]
{\ Histogram of morphological types for the 61 galaxies in our ISOCAM sample.}
\label{fig:morphs}
\end{figure}

The ISOCAM sample of galaxies for the Key Project is presented in Table \ref{tab:camsample} where we list each galaxy's position, optical size, morphology, recession velocity, distance estimate, and far-infrared-to-blue ratio.  The positions, optical sizes, and velocities come from the NASA Extragalactic Database\footnote{The NASA/IPAC Extragalactic Database is operated by the Jet Propulsion Laboratory, California Institute of Technology, under contract with the National Aeronautics and Space Administration.} and were last updated February 2000.  The distances were computed in the Local Group reference frame assuming a Hubble constant of 75 km s$^{-1}$ Mpc$^{-1}$, except for IC 10 for which we rely on results from recent Cepheid variable observations (Wilson et al. 1996; Saha et al. 1996), and for NGC 1569 in which case a distance has been computed from {\it Hubble Space Telescope} observations of resolved supergiants (O'Connell, Gallagher \& Hunter 1994).  The mean distance of the sample is 34 Mpc.  The RC3 (de Vaucouleurs et al. 1991) optical morphological classification for each galaxy was kindly re-examined by H. Corwin using the POSS plates and more recent CCD observations by the Key Project team and those found in the literature; we have updated a total of 17 RC3 morphologies.  Figure \ref{fig:morphs} shows the histogram of optical morphological types.  The morphologies are lenticular or later except for one elliptical, NGC 6958.  There is an approximately even distribution of spirals and a large number of irregulars in the final sample.

\begin{figure}
\centerline{\psfig{figure=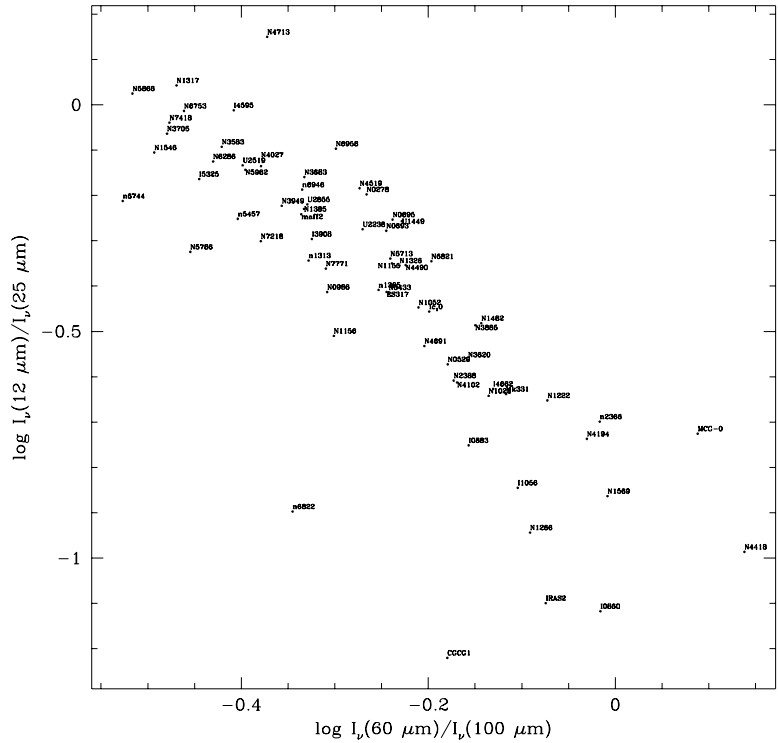,width=4in,bbllx=25pt,bblly=165pt,bburx=565pt,bbury=689pt}}
\caption{\IRAS\ color-color diagram for the 61 galaxies in the U.S. Key Project on normal galaxies ISOCAM sample.}
\label{fig:iras-iras}
\end{figure}

In all, five of the nine relatively extended Key Project galaxies and 56 of the 60 smaller Key Project galaxies were mapped with ISOCAM.  These galaxies span a broad range in physical properties such as total luminosity, dust temperature, and star formation activity.  Figure \ref{fig:iras-iras} is a color-color diagram of ratios of \IRAS\ flux densities for the sample (see Table \ref{tab:fluxes}).  Quiescent galaxies such as NGC 7418 lie along the upper left of the distribution while more active star forming galaxies such as NGC 1569 are found to the lower right.  

\section{The ISOCAM Observations}
Observations were made using the 32$\times$32 pixel Si:Ga long wavelength (LW) array of ISOCAM (C\'{e}sarsky et al. 1996) aboard ISO (Kessler et al. 1996).  We utilized the narrow-band LW1 filter (4.5 \m, $\delta \lambda = 1.0$ \m), the broad-band LW2 filter (6.75 \m, $\delta \lambda = 3.5$ \m), and broad-band LW3 filter (15.0 \m, $\delta \lambda = 6.0$ \m) to sample the mid-infrared emission from the galaxies.  The mid-infrared regime represents a transition from stellar to interstellar emission, though the former should be negligible at 6.75 \m\ for the majority of our sample.  From ISOPHOT observations (Helou et al. 2000) we know there are strong aromatic emission features at 6.2, 7.7 and 8.6 \m, all of which fall within the 6.75 \m\ filter.  The 15 \m\ filter contains some emission from aromatics at 11.3 and 12.7 \m, and continuum emission from very small grains.  We note that strong \NeII\ (12.8 \m) and \NeIII\ (15.6 \m) fine structure lines have been observed in the spectra of \HII\ regions (e.g. M17; C\'esarsky et al. 1996).  Rotational H$_{2}$ lines are also seen in the LW3 bandpass (Kunze et al. 1996, Valentijn et al. 1996, Timmerman et al. 1996), and while these lines are important physically, their contribution to the total flux is very small (Helou et al. 2000).

All of our ISOCAM observations used a gain setting of 2 and a 6\arcsec$\times$6\arcsec\ pixel scale.  Tables \ref{tab:small_maps} and \ref{tab:large_maps} list the date, revolution (orbit number during the \ISO\ mission), and exposure times for each galaxy.  For six galaxies (NGC~1313, NGC~3620, NGC~5713, IC~4595, NGC~6753, NGC~7771), the 4.5 \m\ observations were not done at the same time as the 6.75 and 15 \m\ observations, so the 4.5 \m\ maps for these galaxies are at a different roll angle compared to the 6.75 and 15 \m\ maps.

The bulk of our ISOCAM observations are in the form of 2$\times$2 overlapping raster maps (Table \ref{tab:small_maps}) with a raster step of 75\arcsec, or 12.5 pixel widths, executed along the axes of the array.  Thus the final maps are 4\farcm45$\times$4\farcm45, and the half-integer pixel raster spacing yields an inner 1\farcm95$\times$1\farcm95 portion that is sampled every 3\arcsec; the final maps are rebinned onto a 3\arcsec\ pixel$^{-1}$ grid.  Five galaxies have larger raster maps (Table \ref{tab:large_maps}), with up to 8$\times$8 pointings and a step size of 81\arcsec, or 13.5 pixel widths.  These larger maps are also rebinned onto a 3\arcsec\ pixel$^{-1}$ grid.

\section{Data Reduction}
The ISOCAM observations were reduced using the CAM Interactive Analysis (CIA) software package (Delaney 1998) using Pipeline 7 datasets.  The reduction steps include dark subtraction, deglitching, transient removal, and flat-field correction.  We experimented with various methods for each step, with the goal of finding a reduction method that worked well for the entire galaxy sample.

The arrival of the first wave of ISOCAM observations in mid-1996 allowed us to explore in some detail various reduction and analysis techniques.  This early work indicated that the library dark distributed with each dataset did an acceptable job in removing the effects of the dark current.  The default deglitcher, which uses the multi-resolution median transform method (Starck et al. 1999), did the best job in removing the transitory cosmic ray and charged particle hits without simultaneously flagging the source data as bad data.  Library sky flats from some revolutions have cosmic-ray/charged particle residuals, so we investigated how to use the observations themselves to create a target field sky flat.  We found that as long as the galaxy is not too extended or too bright, we could create a good sky flat from the target field data.  The only remaining task was to remove the effects of the finite response time, which can artificially increase or decrease flux detections, referred to below as ``transient removal.''

All transient removal techniques that existed within CIA circa 1997 were tested to identify the one best suited for our type of observations.  In particular, we studied in detail the relative merits of the `fit3,' `inverse,' and `fouks-schubert' methods.  All other options---dark, deglitcher, target field sky flat---were unchanged in the tests.  Individual pixels from both on and off-source observations were examined before and after transient removal, and the final maps were inspected.  The `fit3' (Siebenmorgen et al. 1997) method appeared to do the best job but it leaves the first raster position, the one most affected by the transient, at a slightly different flux level than the remaining raster positions.  This is because `fit3' removes the transient effect independently for each raster position.  Thus, the average map from each raster position will have a slightly different sky value, with the first raster position being more discrepant (usually by more than 5\%) compared to the remaining raster positions.  To solve this problem, we experimented with masking out the first raster position.  Several of the transient removal methods were re-tested, and again the `fit3' method appeared to work best in terms of removing the transient from individual pixels, both on and off-source, and producing a ``good'' final map.

In summary, the data reduction steps we use were as follows.  First, starting with the Standard Processed Data (SPD) of the pipeline processing, a library dark is subtracted from each frame.  Next, each frame is deglitched using the multi-resolution median transform method.  Then all of the frames in the first raster position are masked out (not used to produce the final map).  Next, transients are removed using the Saclay `fit3' model.  The frames from each raster position are averaged together, and then these average images are used to make a target field sky flat.  For the 4.5 \m\ maps the library sky flat was used, as the sky flux in these maps is too low to produce good target field sky flats from the data.  Next, the flat-fielded average images are mosaicked to create the final map.  The final step converts the pixel units from ADU/gain/second to mJy.


\section{The ISOCAM Images}

\subsection{The Small Maps}
\label{sec:cam_images}
Figure \ref{fig:cam_images} presents optical Digitized Sky Survey (DSS) images and ISOCAM maps for the 2$\times$2 raster sample of galaxies, arranged one row per galaxy.  Figures \ref{fig:ic10}-\ref{fig:n6946} present the larger raster maps, also including the optical DSS images for comparison.  As a consequence of the `dead' column \#24 in the ISOCAM LW array, all of the maps contain two zero value columns towards their right edge.  There is also a notch in the lower left corner of the ISOCAM maps, due to our masking of the first raster position.  In many cases the maps still show low level residuals due to transients from the sources.  In particular, residuals just below the source are common and result from the source being in that position during the third raster observation.  There are also weak residuals in some maps from the second raster position (the upper left in the ISOCAM maps).  These residuals are due to our scheme of masking the first raster position, as they are very weak or non-existent in maps made with all four raster positions.  Such low level residuals are masked out before source fluxes are measured.

NGC 4490 and UGC~2855 are relatively large galaxies compared to the 4\farcm45$\times$4\farcm45 raster maps; for both galaxies the amount of sky area in the final maps is insufficient.  We tried using sky flats from the observations themselves and the library sky flats.  The results using the library sky flats are superior to the target field sky flats, leaving very weak transient residuals and a much smoother sky.  The library sky flats do not appear to add any artifacts to the final maps (as was the case for a few of our galaxies during the testing of reduction methods), and so for these two galaxies fluxes are determined from ISOCAM maps reduced using library sky flats.  We note that the ISOCAM map of NGC 4490 does not contain the entire galaxy; our 6.75 and 15 \m\ fluxes should be considered lower limits for this galaxy.

We successfully detected all galaxies at all mid-infrared wavelengths attempted in our ISOCAM program, (though just barely for two nearby galaxies; see \S \ref{sec:pathetic}).  Overall the ISOCAM maps show many of the same features present in the optical images; prominent spiral arms or rings, and distinct clumps and regions of relatively high mid-infrared emission are quite evident in some of the later type galaxies such as IC 4595, NGC 1385, NGC 3705, NGC 4713, NGC 7418, and UGC 2855.  However, in most cases the bulges seen in the optical are not apparent in the mid-infrared maps.  More importantly, the mid-infrared morphologies at 4.5, 6.75, and 15 \m\ are qualitatively similar.  There are no obvious differences in how the light is distributed, though upon closer inspection of the central regions of some galaxies there are hints of variations with wavelength.  A quantitative look at the relative distribution of mid-infrared emission at 6.75 and 15 \m\ will be presented in Section \ref{sec:curve_of_growth}.  The typical sky background levels are 1, 5, and 20 MJy sr$^{-1}$, with rms fluctuations at the 0.1, 0.1, and 0.2 MJy sr$^{-1}$ levels, at 4.5, 6.75, and 15 \m, respectively.

\subsection{The Larger Maps}
\subsubsection{IC 10} 
IC 10 was observed twice.  In the first set of observations the center of the galaxy was chosen to be the center of the map, as with the maps for all our other galaxies.  However, we noticed that the mid-infrared emission from IC 10 extended more to the west of center, leaving some interesting structure on the edges of the maps.  A second set of observations was taken a year later to more properly map the mid-infrared emission.  The final set of 6.75 and 15 \m\ maps results from a co-add of the maps from the two epochs.  The 11.4 \m\ (LW8 filter) and 4.5 \m\ maps were obtained only in the second epoch.

In the mid-infrared IC 10 appears to consist of a collection of bright knots, except at 4.5 \m\ where the galaxy appears to be quite faint.  IC 10 is classified as a barred irregular galaxy but the bar is not apparent in the mid-infrared maps.  There are two wispy arms extending to the north which contain distinct knots of stronger mid-infrared emission.  An overall faint and diffuse mid-infrared emission is detected towards the west and south of the galaxy center, extending almost to the edge of our maps.  A surface brightness analysis of the 6.75 and 15 \m\ ISOCAM observations of IC 10 can be found in Dale et al. (1999).

\subsubsection{NGC 1313} 
Observations for this galaxy were taken in two parts, a western piece and an eastern piece, with the two pieces mosaicked together to make a final map.  The 4.5 \m\ observation was taken 5 months later at a different roll angle.  In the mid-infrared, the bar of the galaxy appears as discrete emission sources.  The spiral arms are prominent as both diffuse emission and knots of strong emission.  Similar to what is seen optically, there is also a faint wisp of emission directly to the south of galaxy center.  The bright knots of optical emission in the eastern arm also show strongly in the mid-infrared.  Our ISOCAM map is unfortunately too small to cover the large star-forming complex just to the southwest of the galaxy (Ryder et al. 1995).  The galaxy is barely detected at 4.5 \m.  A surface brightness analysis of the 6.75 and 15 \m\ observations of NGC 1313 can be found in Dale et al. (1999).

\subsubsection{NGC 2366 and NGC 6822}
\label{sec:pathetic}
NGC 2366 and NGC 6822 are low luminosity irregulars that contain several moderate to supergiant-sized \HII\ regions (Hunter et al. 2000).  The mid-infrared maps for these galaxies are of insufficient depth: the 1$\sigma$ rms fluctuations in the sky level for NGC 2366 are 0.05 and 0.14 MJy sr$^{-1}$ at 6.75 and 15 \m, respectively, while the numbers for NGC 6822 are 0.09 and 0.28 MJy sr$^{-1}$.  Very little 6.75 or 15 \m\ emission is detected from these two galaxies, suggesting they contain only a small amount of dust.  In NGC 2366 most of the mid-infrared emission coincides with the supergiant \HII\ region NGC 2363 in the southwestern part of the galaxy.  There is a much fainter region of emission to the northeast of that which does not obviously correspond to anything in particular in the optical.  In NGC 6822 the mid-infrared emissison is found in four \HII\ regions.  There is a bit of diffuse emission towards the center of the galaxy, but it is of too low signal-to-noise to measure with any degree of confidence.  The fluxes given in Table \ref{tab:fluxes} are the combined fluxes of the \HII\ regions; we caution that these are only estimates of the total flux, as the true sizes of the galaxies in the mid-infrared are difficult to determine.  There are a few foreground stars seen in the 6.75 \m\ maps.

\subsubsection{NGC 6946} 
At optical wavelengths, NGC 6946 is a striking example of a nearly face-on ($i$\about 30$^\circ$) spiral galaxy.  The galaxy is equally rich in detail in the mid-infrared.  The spiral arms can easily be traced in the mid-infrared maps, and each contains several knots of brighter emission.  The `sharp' edge of the northeastern-most arm is real and not an artifact of the mosaicking.  It is also seen in \IRAS\ HiReS maps of NGC 6946.  Preliminary work on NGC~6946 from these ISOCAM observations (using earlier data reduction techniques) by Helou et al. (1996) indicated that there is very little mid-infrared color variation throughout the disk of the galaxy.  This result is echoed by the work of Tuffs et al. (1996), who used ISOPHOT data to suggest that the bulk of the far-infrared luminosity is due to a uniformly colored and rather cold diffuse emission from the disk.  Work by Malhotra et al. (1996), again using these same ISOCAM observations reduced with prior techniques, showed that the mid-infrared emission is consistent with an exponential disk of scale length of 75\arcsec, based on a flux profile computed from the median flux within annuli between 70 and 200\arcsec.  They also indicate that the mid-infrared arm-interarm contrast, being close to that observed in $R$-band but lower than observed in H$\alpha$, suggests that non-ionizing radiation plays an important role in the heating of the dust.  A surface brightness analysis of the ISOCAM observations for NGC 6946, using our new reduction methods, can be found in Dale et al. (1999).  We stress that the main results do not change: the color variations analyzed in Dale et al. are quite close to those presented earlier.  In fact, the small discrepancies (\about 10\%) between the two analyses can be taken as a first-order estimate of the uncertainties in the data.  Moreover, using the newly-processed data we find a similar mid-infrared disk scale length (72\arcsec).  Finally, we point out that the starbursting center is quite bright in the mid-infrared; our integration times nearly saturated two pixels at 15 \m\ and saturated as many as four pixels at 6.75 \m.  The impact of these saturated and nearly saturated pixels on the overall flux density is measurably small, approximately a 15\% (25\%) effect at 15 (6.75) \m\ for affected pixels (the method used to approximate this effect is described in Section \ref{sec:mid-ir_flux}). Our values for the total flux density should be formally taken as lower limits, though the underestimation is no more than a few percent. 

\section{Flux Density Estimates}
Before beginning any quantitative analysis on the ISOCAM images presented in Figures \ref{fig:cam_images}-\ref{fig:n6946}, we cleaned the images by masking any foreground stars and cosmic ray/transient residuals.  To allow direct comparison between the various filters we smoothed the maps to a similar resolution.  To do this we needed to first estimate the approximate resolution for the 6.75 and 15 \m\ maps.    There are a few maps which have strong emission from a foreground star (e.g. NGC 1569 and NGC 6946), and several galaxies appear to be unresolved at these mid-infrared wavelengths (\IRAS\ F10565, NGC 4418, IC 860, IC 883, CGCG1510.8+0725, \IRAS\ F23365+3604, Markarian 331).  Analysis of these images indicate a (FWHM) resolution of \about 7\arcsec\ and \about 8\arcsec\ at 6.75 at 15 \m, respectively.  This information was used to convolve the 6.75 and 15 \m\ maps, using slightly different Gaussian smoothing profiles for each filter, to generate maps with approximately uniform resolution (9\arcsec\ FWHM).  As a final step the images were registered to the coordinate system of the 15 \m\ map by assuming the peak emission at 6.75 \m\ is spatially coincident with the peak emission at 15 \m.

\subsection{Mid-Infrared Flux Densities}
\label{sec:mid-ir_flux}
Multi-aperture photometry is our preferred method for extracting mid-infrared flux densities.  For each galaxy care is taken to define the apertures and background annulus using the flux density distribution from both the 6.75 and 15 \m\ (and when available the 4.5 \m) maps.  In each case, a series of evenly spaced, concentric apertures are centered on the central emission peak (or as in the case of NGC 1569 which has two central emission peaks, the brightest of the central emission peaks), with the outermost aperture sized according to the greatest extent of the emission in either map.  The background annulus was then defined starting from the outermost aperture, unless a localized contamination (e.g. star) warrants starting the background annulus at a slightly larger radius; the typical background annulus has a width of 15\arcsec.  The mean value within the background annulus is taken to be the sky background level.  As described in Section \ref{sec:cam_images}, the typical rms fluctuation in the background level is 0.1--0.2 MJy sr$^{-1}$.  The background-subtracted flux density within the outermost aperture is the total flux density.

The mid-infrared flux densities and their uncertainties for the galaxies are given in Table \ref{tab:fluxes} along with the \IRAS\ flux densities.  Column 1 in Table \ref{tab:fluxes} lists the galaxy.  Columns 2--5 list the \IRAS\ flux densities in Janskys and columns 6--7 list the \ISO\ 6.75 and 15 \m\ flux densities and their uncertainties in Jy (see \S \ref{sec:cam-12mu} for a comparison of ISOCAM and \IRAS\ 12 \m\ fluxes).  We assume a 20\% uncertainity in the flux density calibration of ISOCAM data (Blommaert \& Cesarsky 1998; Biviano 1998).  By far, this uncertainty level dominates any other uncertainties that are introduced by our method of flux density extraction (e.g. choice of aperture and background annulus, sky rms fluctuation, etc.).  However, our finite aperture measurements systematically underestimate the total flux density of the galaxies by a small amount due to the limiting \about 0.1 MJy sr$^{-1}$ sensitivity of our observations.  We can estimate the amount of low surface brightness emission that is missed by extrapolating the observed isophotal surface brightness trends.  Assuming such trends hold out to a surface brightness of 0.001 MJy sr$^{-1}$ (about two disk scale-lengths), no more than a few percent of the flux is missing.  Overall, for most galaxies we thus find a total flux density uncertainty of 20\%; this number is slightly higher (\about 25\%) for about one fourth of the sample due to more significant sky background fluctuations.  

The uncertainty is slightly larger in cases where i) the mid-infrared emission extends beyond the observed target field, or ii) there is saturation, two additional effects that systematically lower the flux density measure.  Only the fluxes for UGC 2855 and NGC 4490 suffer from the former effect; extrapolations of the observed profiles, as described above, show we are missing $\lesssim$ 5\% of such flux for these two galaxies.  A second concern is that observations of a few galaxies saturated the analog to digital converter for a small number of pixels, the effect being more serious for the longer elementary integrations for the 6.75 \m\ bandpass.  We have gauged the amplitude of this effect using the known response function of the ISOCAM detectors: the actual flux density should be larger than the initial frame's measure by a factor 1/0.6 (ISOCAM Observer's Manual).  For most of the 11 galaxies exhibiting saturated pixels the net effect is an underestimate of the total flux density by 6\% or less; the 6.75 and 15 \m\ flux densities for NGC 4418 are underestimated by about 15\% and 25\%, respectively, comparable to the calibration uncertainty level.  The flux densities computed for these galaxies are lower limits and are indicated by a colon in Table \ref{tab:fluxes}; they may be assumed to be uncertain at the 25\% level.

Table \ref{tab:4.5mu} lists the total 4.5 \m\ flux densities for 13 galaxies in our sample.  For NGC~1569 we obtained 2$\times$2 ISOCAM maps in eight filters; NGC~1569 flux densities and uncertainties are given in Table \ref{tab:n1569}.  The typical LW1 flux density uncertainty is 25-30\% due to a lower overall signal to noise compared to the LW2 and LW3 maps.

\subsection{Calibration Check: The Consistency of \ISO\ and \IRAS\ Fluxes}
\label{sec:cam-12mu}
\begin{figure}
\centerline{\psfig{figure=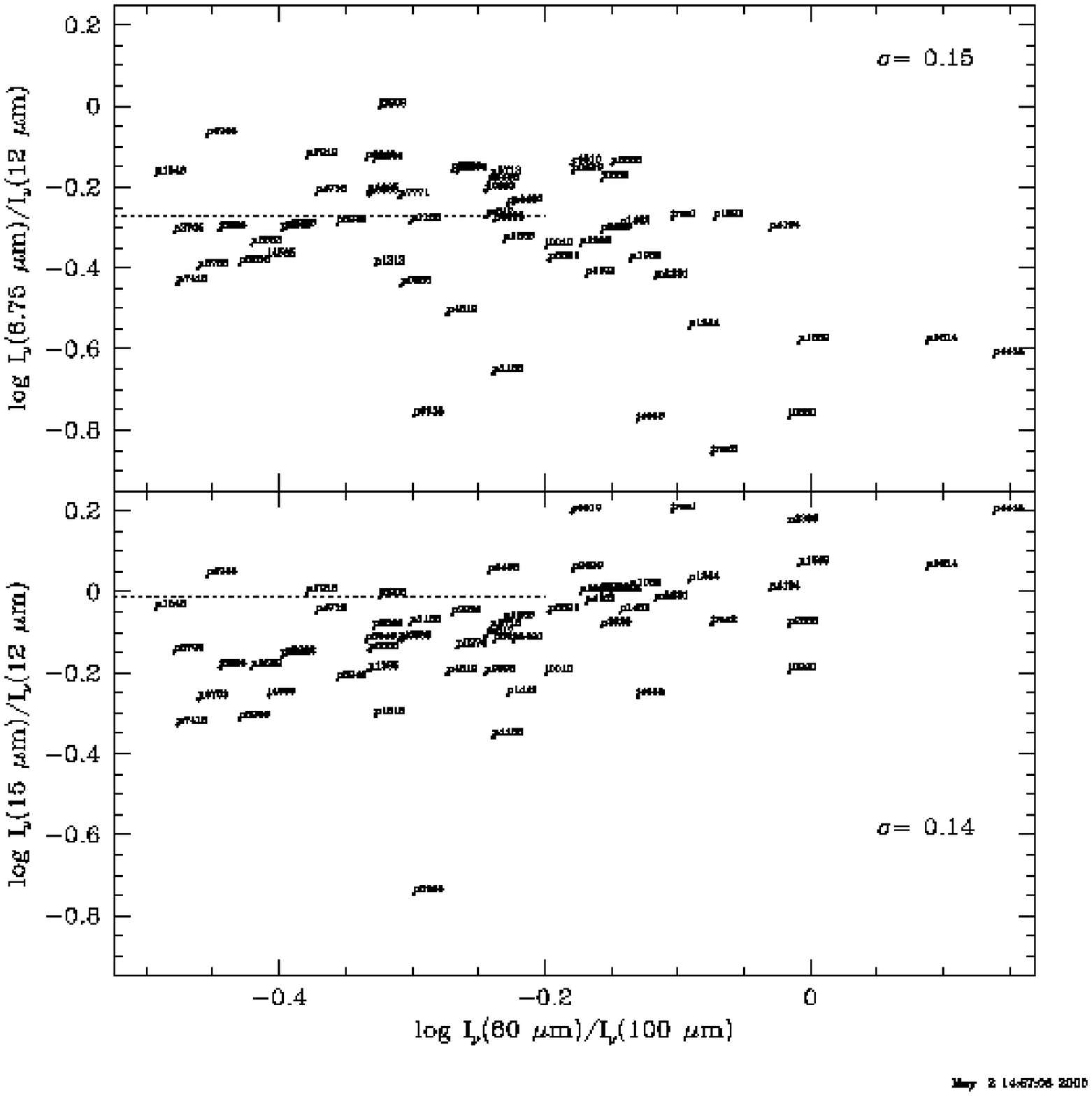,width=5in,bbllx=17pt,bblly=165pt,bburx=565pt,bbury=724pt}}
\caption{Color-color diagrams showing the comparison of \ISO\ fluxes to \IRAS\ 12 \m\ fluxes.  The dashed lines indicate the \ISOIRASa\ and \ISOIRASb\ values for a cirrus model (Reach 1999).  The $\sigma$ values are the standard deviations in the ordinates between $-0.5<$ log \IRAScolor$<-0.2$.}
\label{fig:cam-12mu}
\end{figure}

Lu et al. (2000) show that the agreement between ISOPHOT and ISOCAM LW2 fluxes is better than 20\%.  An \ISO-independent check of the calibration of our ISOCAM maps can be provided by \IRAS\ data.  Figure \ref{fig:cam-12mu} presents the ratio of ISOCAM fluxes to \IRAS\ 12 \m\ fluxes versus \IRAScolor\  color.  The dashed lines indicate the expected ratios for cirrus (log \ISOIRASa\ = $-0.29$ and log \ISOIRASb\ = $-0.01$ for $\rho$ Ophiuchi; W. Reach, private communication).  The expected cirrus color agrees well with our observed \ISOIRASa\ color ratio: log \ISOIRASa [$-0.5<$ log \IRAScolor$<-0.2$]=$-0.26\pm0.11$ (the uncertainty represents the population dispersion excluding NGC 1155 and NGC 6958).  The agreement is even better for the more quiescent galaxies, even though real galaxies have other emission (star-formation regions) in them in addition to cirrus.  Moreover, we expect the \ISOIRASa\ color to decrease for warmer galaxies (relatively high \IRAScolor\ color) since the PAH features in the 6.75 \m\ filter become relatively weak compared to the rising continuum in the 12 \m\ band, and this is indeed observed.  Photospheres can contribute to the flux at 6.75 \m\ and throw off the ratio, especially for measurements of elliptical galaxies.  Since this effect does not seem significant in this sample, we conclude that in most of our sample the 6.75 \m\ and longer wavelength bands are dominated by interstellar emission.  

For the \ISOIRASb\ color ratio comparison with cirrus, the agreement is not as good, but the ratio observed for galaxies is lower than that of cirrus only by 35\%: log \ISOIRASb [$-0.5<$ log \IRAScolor$<-0.2$]=$-0.14\pm0.10$ (excluding NGC 6958).  Surprisingly, the offset goes in the wrong direction for astrophysical effects to resolve the discrepancy: the 15 \m\ band should pick up more emission in galaxies than the cirrus spectrum represents, and we reach the ``cirrus ratio'' only for mildly active galaxies.  There are a few possible explanations for this discrepancy, most importantly the uncertainty in the cirrus spectrum (calibration uncertainty and spatial variations in the cirrus behavior), the uncertainties in the various filter band passes, and incompleteness in the \ISO\ flux integral.  The \ISO\ 15 \m\ maps are integrated only down to the \about 0.2 MJy sr$^{-1}$ isophote, whereas the large angular size of the \IRAS\ detectors tended to pick up contributions from extended diffuse emission.  However, as pointed out in Section \ref{sec:mid-ir_flux}, extrapolations of the observed surface brightness profiles down to 0.001 MJy sr$^{-1}$ show we are missing at most 5\% of the total \ISO\ flux.  Moreover, if the problem is \ISO\ flux incompleteness, we might expect a similar discrepancy in the \ISOIRASa\ comparison; the \ISO\ 6.75 \m\ maps go only slightly deeper, to \about 0.1 MJy sr$^{-1}$.  We conclude that it is unlikely that \ISO\ flux incompleteness is responsible.

\section{Analysis}
\subsection{Mid-Infrared Flux Curve of Growth Profiles}
\label{sec:curve_of_growth}
\begin{figure}
\centerline{\psfig{figure=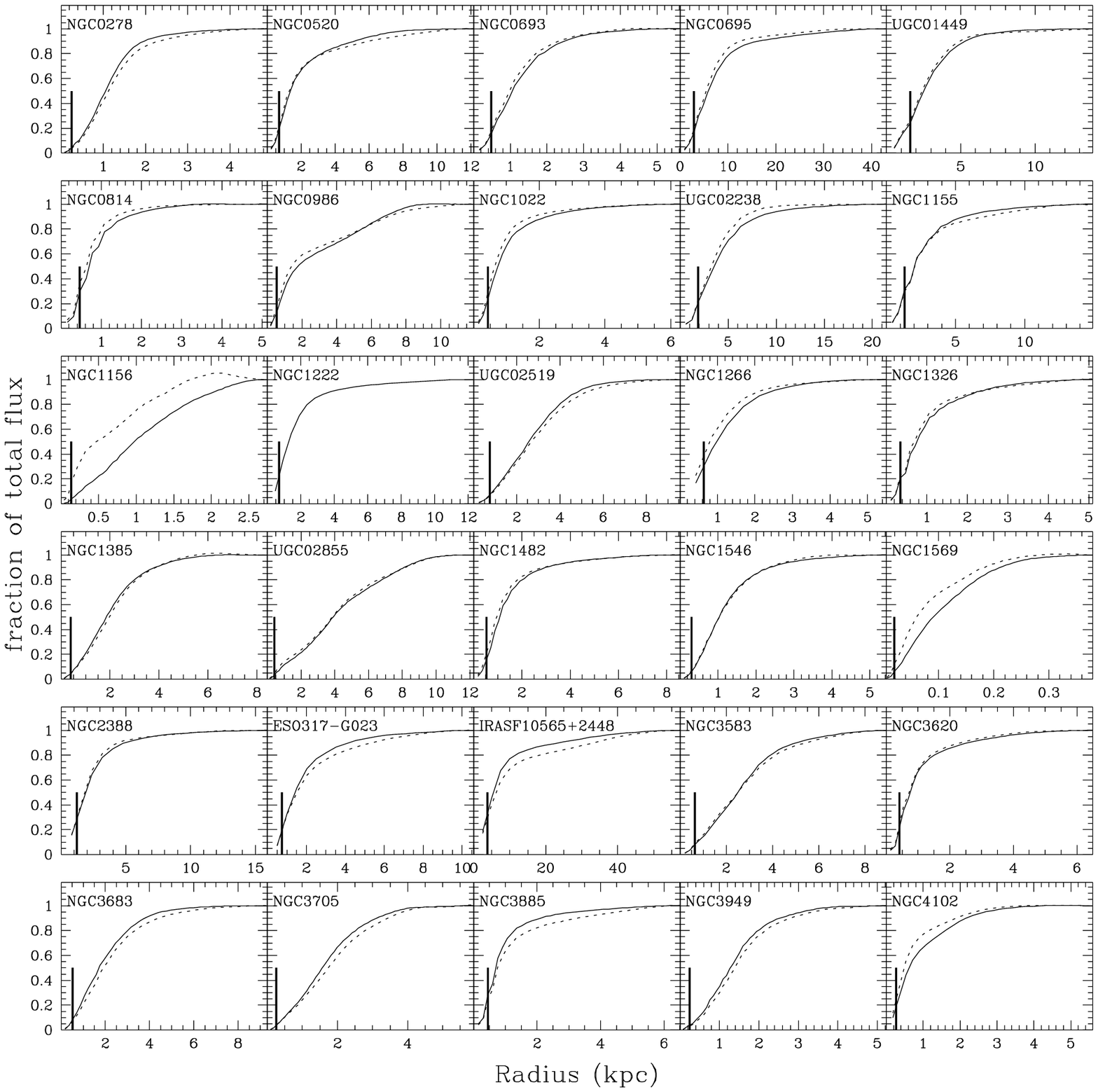,width=4.3in,bbllx=30pt,bblly=155pt,bburx=589pt,bbury=714pt}}
\centerline{\psfig{figure=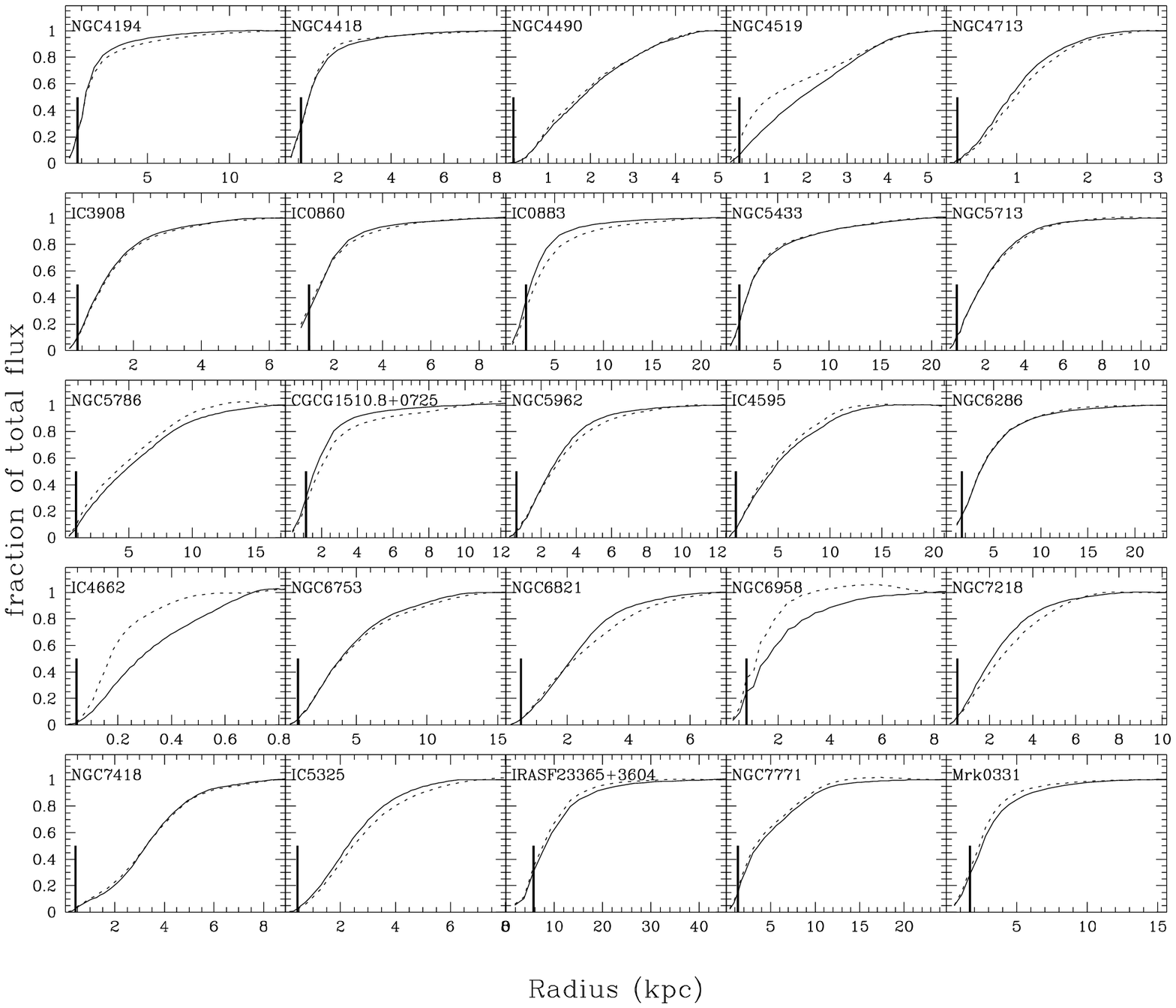,width=4.3in,bbllx=30pt,bblly=231pt,bburx=589pt,bbury=714pt}}
\caption{Mid-infrared flux curve of growth profiles for the galaxies with
2$\times$2 raster maps. The solid line is the growth profile for 6.75 \m.  The 
dashed line is the growth profile for 15 \m.  The short bold solid line on the
abscissa indicates the 4\farcs5 HWHW resolution limit for our images.}
\label{fig:curves_of_growth}
\end{figure}

\begin{figure}
\centerline{\psfig{figure=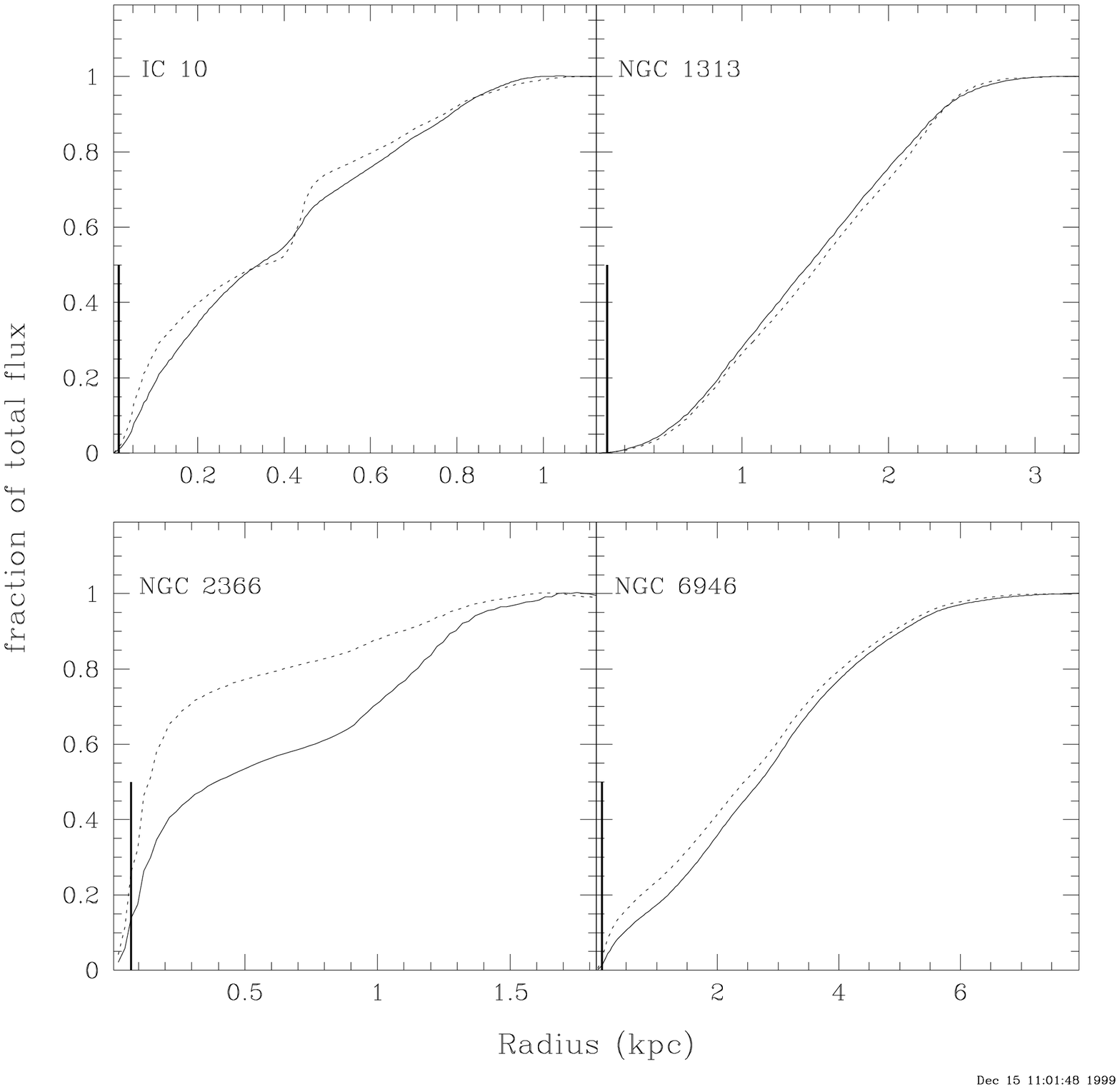,width=4.5in,bbllx=22pt,bblly=156pt,bburx=572pt,bbury=698pt}}
\caption{Mid-infrared flux curve of growth profiles for the galaxies with
relatively large raster maps. The solid line is the growth profile for 6.75 \m.  The dashed line is the growth profile for 15 \m.  The short bold solid line on the abscissa indicates the 4\farcs5 HWHW resolution limit for our images.}
\label{fig:curves_of_growth_nearby}
\end{figure}

Results of the aperture photometry described in Section \ref{sec:mid-ir_flux} are presented in Figures \ref{fig:curves_of_growth} and \ref{fig:curves_of_growth_nearby}.  From the smoothed maps, the curve of growth profiles for the 6.75 and 15 \m\ emission are represented by the solid and dashed lines, respectively.  Each profile is flux-normalized to the total flux observed in the respective filters.  The bold, solid vertical line indicates the resolution scale in physical units and corresponds to the 4\farcs5 HWHM Gaussian smoothing profile.  The emission from most galaxies is very well resolved in the maps:  for about half (45\%) of the galaxies in our sample, 10\% or less of the mid-infrared light arises from within the central ISOCAM beam; one-sixth of the galaxies emit $>$ 25\% in the central beam.  While the physical extent and concentration of the mid-infrared emission varies from galaxy to galaxy, the 6.75 and 15 \m\ profiles are remarkably similar for a given source.

Against this general similarity, some galaxies display strong central \ISOcolorb\ ratios, which can be identified in Figures \ref{fig:curves_of_growth} and \ref{fig:curves_of_growth_nearby} by a narrower 15 \m\ than 7 \m\ curve.  The most notable cases are NGC 1156, NGC 1569, NGC 4519, IC 4662, NGC 6958, and NGC 2366.  As discussed below (Section \ref{sec:iso-iras}), larger values of \ISOcolorb\ point to more intense heating of the dust, suggesting that the galaxies with this central color enhancement harbor high densities of \HII\ regions in their centers, or nuclear starbursts.  This is indeed consistent with the properties of NGC 1156, NGC 1569, IC 4662, and NGC 2366, which are all Magellanic Irregulars with bright star-forming regions near their centers (Hunter et al. 2000; Ho et al. 1995).  It is interesting to note that the reverse situation, namely a decrease in the \ISOcolorb\ color, is not observed in this sample.

\subsection{The \ISO -\IRAS\ Color Diagram}
\label{sec:iso-iras}
\begin{figure}
\centerline{\psfig{figure=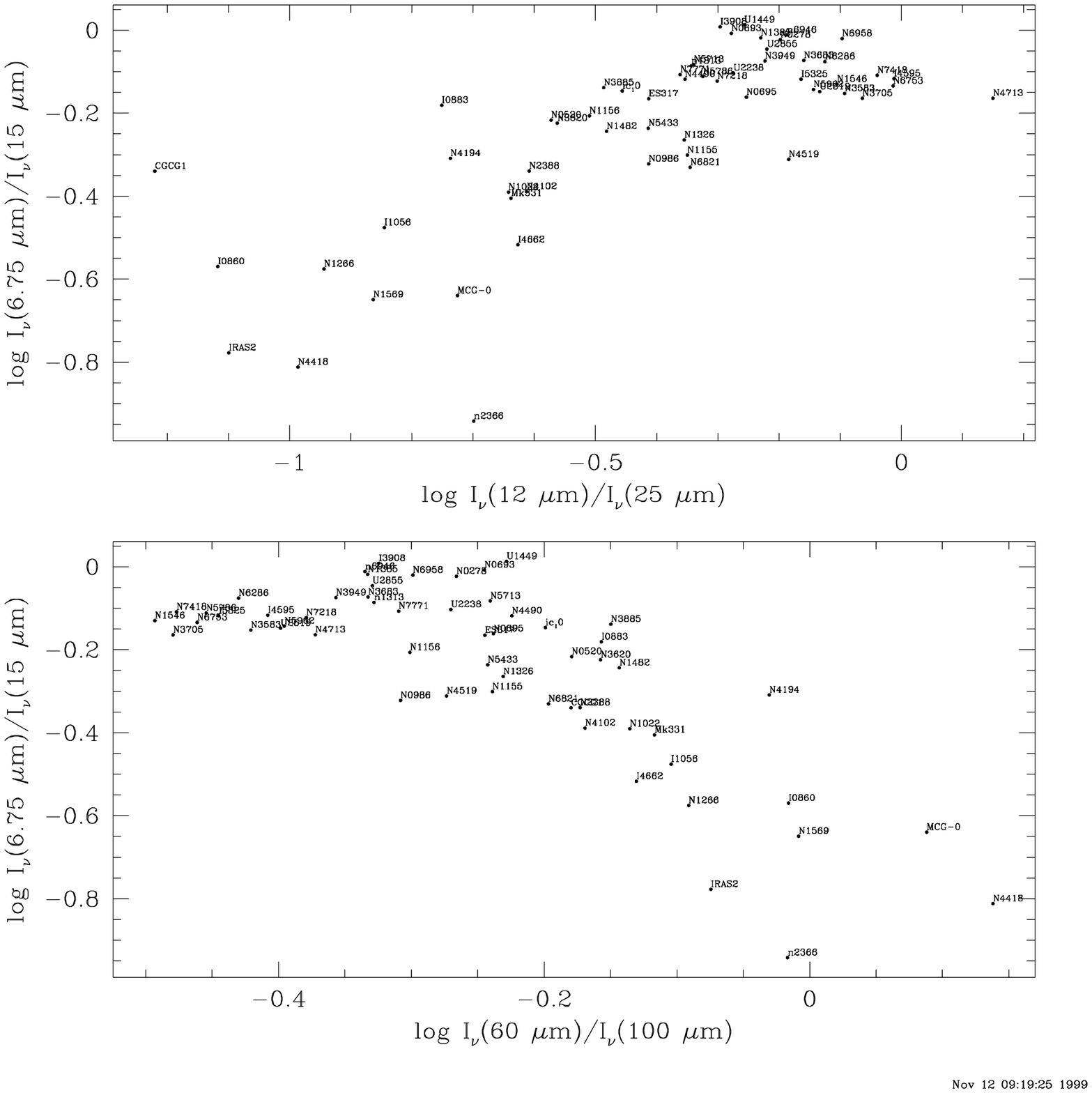,width=5in,bbllx=20pt,bblly=165pt,bburx=565pt,bbury=724pt}}
\caption{Mid-infrared versus far-infrared color-color diagrams.  For a range of moderate heating intensity levels, $-0.5<$ log \IRAScolor$<-0.2$, normal galaxies have a similar mid-infrared color.}
\label{fig:iso-iras}
\end{figure}

The \ISOcolor\ ratio has emerged as an interesting diagnostic of the radiation environment (Helou et al. 1997).  As evident in the \ISO-\IRAS\ color-color diagram (Figure \ref{fig:iso-iras}), this ratio remains relatively constant and near unity as the interstellar medium of galaxies proceeds from quiescent to mildly active, where the level of activity is indicated by a rising \IRAScolor\ (Helou 1986).  As dust heating increases further, the flux at 15 \m\ increases steeply compared to 7 \m.  The data plotted in Figure \ref{fig:iso-iras} are consistent with an inflection in the mean trend occurring near log \IRAScolor =$-0.2$, which we interpret as due to excess emission in the 15 \m\ band rather than a drop in the 7 \m\ band.  The main argument for this interpretation is that the $I_\nu(6.75 \mu$m)/FIR ratio does not drop as precipitously as \ISOcolor\ for these objects.  We assign to this emission a characteristic temperature 100 K $< T_{\rm MIR} <$ 200 K, since that is the range that would allow a blackbody to contribute considerably to the 15 \m\ band but not to the 7 \m\ band; the estimates hold for modified blackbodies as well.  Such values of $T_{\rm MIR}$ are typical of heating intensities about $10^4$ times greater than the diffuse interstellar radiation field in the Solar Neighboorhood (Helou et al. 1997).  While such a temperature could result from classical dust heated within or just outside \HII\ regions, there is no decisive evidence as to the size or location of grains involved or whether in fact they are in equilibrium with the radiation field.  It is simpler at this time to associate this
component empirically with the observed emission spectrum of \HII\ regions and their immediate surroundings (Tran 1998; Contursi et al. 2000).  This emission has severely depressed aromatic feature or PAH emission, and is dominated by a steeply rising though not quite a blackbody continuum near 15 \m, consistent with mild fluctuations in grain temperatures, $\Delta T/T \sim 0.5$.  This \HII\ region hot dust component characterized by color temperatures 100 K $< T_{\rm MIR} <$ 200 K becomes detectable in systems where the color temperature from the \IRAScolor\ ratio is only $T_{\rm FIR} \approx 50$ K (Helou et al. 1988).  This disparity between color temperatures derived from different wavelengths demonstrates the broad distribution of dust temperatures within any galaxy.\footnote{The quoted far-infrared color temperatures are merely approximations arising from graybody profiles that exhibit similar flux ratios for the two wavelength bands.  See Helou et al. (1988) for more details on \IRAScolor\ color temperatures.}  The combined data from \ISO\ and \IRAS\ on these systems are consistent with an extension of the ``two-component model'' of infrared emission (Helou 1986).  The low \ISOcolor\ ratio is associated with the active component, and combines in a variable proportion with a component with a \ISOcolor\ near unity (Dale et al. 1999). 

\subsection{Mid-Infrared vs Optical Sizes}
\begin{figure}
\centerline{\psfig{figure=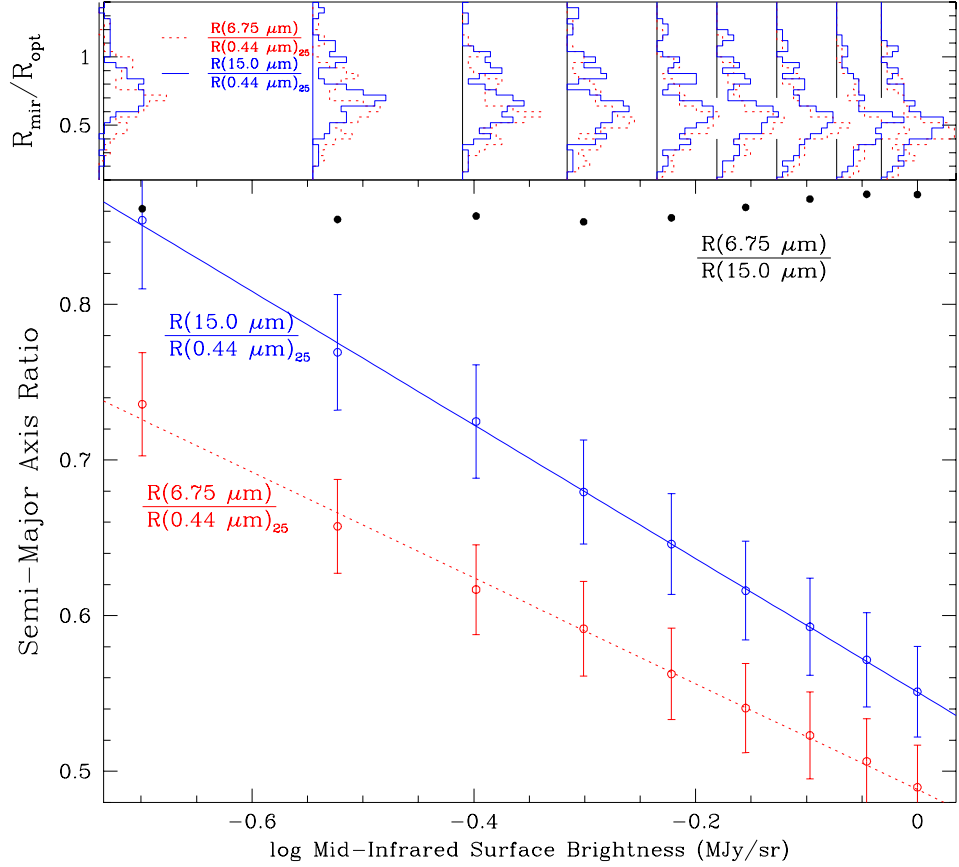,width=5in,bbllx=23pt,bblly=175pt,bburx=584pt,bbury=690pt}}
\caption{The size of the mid-infrared semi-major axis, normalized to the $B$-band 25 mag arcsec$^{-2}$ semi-major axis $R(0.44 \mu{\rm m})_{25}$, as a function of the infrared surface brightness level.  The mid-infrared size at 6.75 \m\ (15 \m) matches $R(0.44 \mu{\rm m})_{25}$ at a surface brightness level of 0.04 MJy sr$^{-1}$ (0.09 MJy sr$^{-1}$).  The histograms at the top of the diagram indicate the distributions of size ratios at each mid-infrared surface brightness level.  The filled circles show the running ratio of the 6.75 to 15 \m\ sizes.}
\label{fig:opt-ir_sizes}
\end{figure}

It is interesting to measure the size of a galaxy in the mid-infrared in comparison to an equivalent measure at optical wavelengths, e.g. $R(0.44 \mu{\rm m})_{25}$, the length of the semi-major axis out to 25 mag arcsec$^{-2}$ in $B$ (obtained from the RC3 catalog; de Vaucouleurs et al. 1991).  Using fits of isophotal ellipses to the mid-infrared distribution, we measure semi-major axis lengths from both the 6.75 and 15 \m\ maps at several surface brightness levels between 0.2 and 1.0 MJy sr$^{-1}$.  We fit elliptical isophotes to the mid-infrared images using both standard and customized IRAF packages (see Haynes et al. 1999 for further details on the GALPHOT surface photometry fitting routines).  We do not measure sizes at lower surface brightness levels due to S/N constraints, since as discussed in Section \ref{sec:cam_images}, the typical rms fluctuation in the sky background is $0.1-0.2$ MJy sr$^{-1}$.  
The mid-infrared semi-major axes are corrected for ISOCAM resolution using a quadratic formula $R=\sqrt{R_{obs}^2-4\farcs5^2}$.  The open circles in Figure \ref{fig:opt-ir_sizes} show the mean \Rmira/\RB\ and mean \Rmirb/\RB\ for the sample as a function of the mid-infrared surface brightness level at which mid-infrared sizes are measured.  Error bars reflect the one sigma population dispersions in the ratios at each surface brightness, divided by the square root of the number of galaxies; histograms of the ratios at each surface brightness level are also displayed at the top of the figure.  The data are reasonably well approximated by an exponential dependence of mid-infrared surface brightness on radius, and the trend holds for the entire range of measured surface brightness levels.  Invoking a small extrapolation of the trend to lower surface brightness levels shows that, on average, mid-infrared sizes at 6.75 and 15 \m\ match \RB\ at a surface brightness level of $I_\nu(6.75 \mu{\rm m}) \approx 0.04$ MJy sr$^{-1}$ and $I_\nu(15 \mu{\rm m}) \approx 0.09$ MJy sr$^{-1}$ respectively.  The uncertainties in these surface brightness estimates are 25\%.  


Since \RB\ is comparable to the location of the Sun in the Milky Way, one would expect similar values of the mid-infrared surface brightness of the Milky Way in the Local Neighborhood.  For a typical high Galactic latitude \HI\ column density of $\sim 2-3 \times 10^{20}$ cm$^{-2}$ (Kulkarni \& Heiles 1988), and for an emissivity of $4\pi \nu I_\nu(12 \mu{\rm m}) = 1.1 \times 10^{-31}$ W per H atom (Boulanger \& P\'{e}rault 1988), the expected mid-infrared brightness would be $I_\nu(12 \mu{\rm m}) \sim 0.08$ MJy sr$^{-1}$, indeed comparable to the values above.  In addition, one can translate the 12 \m\ band flux density into expected 6.75 and 15 \m\ flux densities using the cirrus spectrum of Reach and Boulanger (1998).  One finds $I_\nu(6.75 \mu{\rm m}) \sim 0.04$ MJy sr$^{-1}$ and $I_\nu(15 \mu{\rm m}) \sim 0.09$ MJy sr$^{-1}$, the same numbers we find for the Key Project sample.  In short, the typical mid-infrared emission from the outskirts of normal galaxies is reassuringly similar to what we see in the quiescent regions of the Milky Way.

We have compared galaxy mid-infrared sizes to their optical sizes, in particular those defined at the 25 $B$-mag arcsec$^{-2}$ level.  Since this surface brightness corresponds to 0.018 MJy sr$^{-1}$ (assuming $f_B=4260 \times 10^{-0.4 m_B}$ Jy), the observed mid-infrared to $B$-band flux ratios at \RB\ are
\begin{equation}
{\nu I_\nu(6.75 \mu{\rm m}) \over \nu I_\nu(0.44 \mu{\rm m})_{25}} \simeq0.12\pm0.03 
\;\;\;\;\;\;\;\;\;
{\nu I_\nu(15 \mu{\rm m}) \over \nu I_\nu(0.44 \mu{\rm m})_{25}} \simeq0.15\pm0.04
\end{equation}
In other words, approximately eight (seven) times the flux is emitted at the center of the $B$-band as compared to the emission at 6.75 \m\ (15 \m), in the outskirts of normal galaxies.  Once again we can contrast this with observations of Galactic cirrus.  De Vaucouleurs \& Pence (1978) estimated the $B$-band surface brightness of the Galactic disk towards the Galactic poles to be $\mu_B = 24.92$ mag arcsec$^{-2}$.  Thus we infer a ratio for the Galactic cirrus to be approximately
\begin{equation}
{\nu I_\nu(12 \mu{\rm m}) \over \nu I_\nu(0.44 \mu{\rm m})}_{\rm Gal. \; cirrus} \simeq0.15,
\end{equation}
comparable to our results at 6.75 and 15 \m.  

Alternatively, it may be more illuminating to compare for quiescent regions the integrated total-infrared flux to the integrated flux in the $B$-band.  From this comparison we can infer the ratio of heating output to heating input (i.e. the total-infrared flux to the UV and optical energy responsible for the dust heating).  As a start, we contrast the integrated flux within the LW2 and LW3 bands to that in the $B$-band ($\Delta \nu/\nu \approx 0.22$):
\begin{equation}
{\Delta \nu I_\nu({\rm 6.75}) + \Delta \nu I_\nu({\rm 15}) \over \Delta \nu I_\nu(0.44 \mu{\rm m})_{25}} \simeq 0.53\pm0.2.
\label{eq:lw2lw3B}
\end{equation}
Note that this is only valid for the outer regions of normal star-forming galaxies.  To infer the total-infrared emission for diffuse cirrus regions using only the flux in the LW2 and LW3 bands, we have to rely on a model for the infrared spectral energy distribution of galaxies.  Dale et al. (2000) present such a model for normal star-forming galaxies.  Constrained by IRAS, ISOCAM, and ISOPHOT observations for our sample of 69 normal galaxies, the model reproduces well the empirical spectra and infrared color trends.  It also allows us to determine the infrared energy budget for normal galaxies, and of particular interest here, to translate mid-infrared fluxes into total-infrared fluxes.  The model shows that, for cirrus-dominated regions, the total-infrared emission (from 3--1100 \m) is approximately 10.3 times the emission appearing within just the LW2 and LW3 mid-infrared bands (and approximately 2.8 times the far-infrared emission from 42--122 \m).  Coupling this result with that expressed by Equation \ref{eq:lw2lw3B}, we see that 
\be
{{\rm TIR(3-1100 \mu m) \over \Delta \nu I_\nu(0.44 \mu {\rm m})_{25}}} \sim 5.4
\label{eq:fir-b}
\ee
in the outskirts of normal galaxies.  Observations of cirrus-dominated galaxies show that this finding is representative of quiescent disks: in our sample of normal late-type galaxies, the lowest such integrated total-infrared to blue ratios, assuming TIR/FIR $\approx 2.8$ for the most quiescent regions, are of order 3.

From the above ratio (Equation \ref{eq:fir-b}), one can derive the ratio of the heating output to the heating input as an indication of the optical depth in the interstellar medium.  This is expected to be significantly less than unity for the most quiescent regions.  All photon energies from the UV to the near-infrared contribute to the heating of dust grains.  In fact, photons responsible for the $B$-band flux only contribute about 10-15\% towards the total far-infrared flux in the outer parts of the disk of M31, a galaxy well-known for its overall quiescent behavior (Xu \& Helou 1996 and C. Xu personal communication).  Thus if we assume the results from the outer disk regions of M31 can reasonably apply for the outermost portions of our normal galaxy sample, the ratio of the total heating output to the total heating input for quiescent regions is typically of order 0.7 (to within a factor of two).

\subsection{The Nature of the Flux at 4.5 \m}
The transition from stellar to interstellar emission is well illustrated by the spectra of Virgo Cluster galaxies collected by Boselli et al. (1998).  Its precise location and therefore the interpretation to attach to mid-infrared fluxes can be parametrized by the ratio of far-infrared to visible light fluxes.  Interstellar dust emission takes over by 5 \m\ when the FIR/B ratio exceeds 0.5, and at shorter wavelengths for higher ratios (see Table \ref{tab:camsample} for the definition of FIR/B).  As might be expected, elliptical galaxies are dominated by stellar emission, both photospheric and from circumstellar dust shells, and therefore provide the templates that one subtracts to isolate the interstellar emission component in spiral galaxies (Boselli et al. 1998; Madden, Vigroux \& Sauvage 1999).  

\begin{figure}
\centerline{\psfig{figure=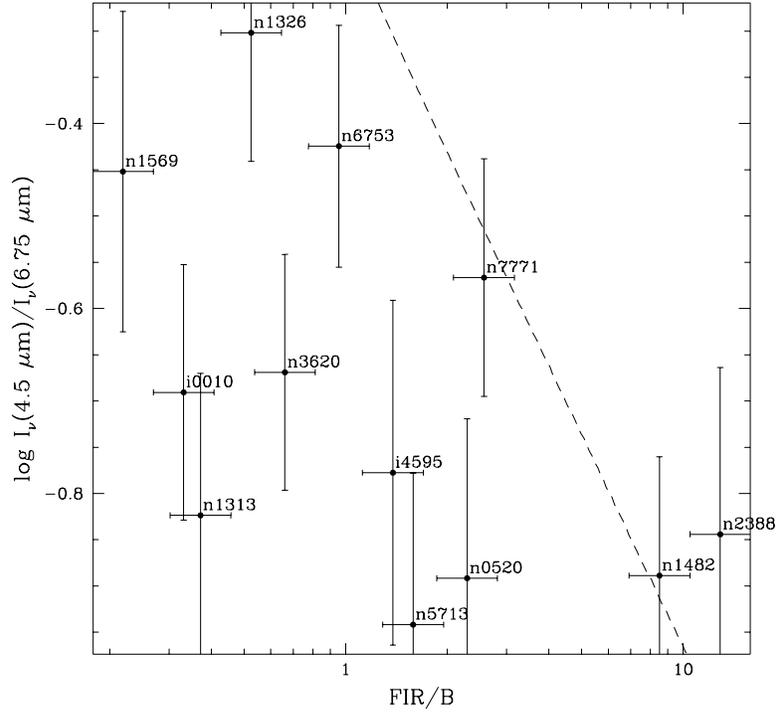,width=4in,bbllx=18pt,bblly=163pt,bburx=580pt,bbury=690pt}}
\caption{A comparison of the 4.5 to 6.75 \m\ emission as a function of the far infrared to blue ratio.  A slight trend is seen in the expected sense (see text), whereby galaxies with a higher FIR/B exhibit comparatively more 4.5 \m\ emission.  The expected 4.5 to 6.75 \m\ logarithmic ratio for photospheric emission (assuming a $T \sim 3500$ K blackbody; see Boselli et al. 1998) is 0.28, consistent with the findings of Madden, Vigroux \& Sauvage (1999) for early type galaxies.}
\label{fig:5to7_vs_IRtoB}
\end{figure}

We do not expect large contributions from photospheric emission at 4.5 \m\ for our sample in light of the results from mid-infrared ISOPHOT spectrophotometry for 45 Key Project galaxies.  The shape and flux level of the 3--5 \m\ continuum shows no variation with the strength of the aromatic features in emission for galaxies with global far infrared to blue ratios greater than unity (and very little variation otherwise), direct evidence that this portion of the continuum arises primarily from fluctuating dust grain emission (Helou et al. 2000; Lu et al. 2000).  We can similarly explore the nature of the 4.5 \m\ emission by analyzing the ratios of broad band fluxes.  Figure \ref{fig:5to7_vs_IRtoB} displays the \ISOcolorc\ color ratio as a function of FIR/B.  Gauging by the range of observed \ISOcolorc\ ratios, we can immediately rule out the possibility that the emission at both 4.5 and 6.75 \m\ is purely photospheric: the typical logarithmic value for the early type galaxy sample of Madden, Vigroux \& Sauvage (1999) is much higher (log \ISOcolorc\ $=$ 0.1--0.4), consistent with the ratio for a $T=3500$ K blackbody profile that has been suggested for the stellar component of normal galaxies (Boselli et al. 1998).  Thus for our sample of normal galaxies, the average \ISOcolorc\ ratio tells us that the 6.75 \m\ emission is definitely dominated by emission from interstellar dust grains, but does not constrain the origin of the 4.5 \m\ emission.

The slight trend in the data can tell us more.  First, the trend is in the sense of a ``photospheric excess'' at 4.5 \m\ for the galaxies with lower FIR/B.  Second, the slope of this trend is consistent with the extreme scenario whereby the 4.5 \m\ emission is proportional to that in the $B$-band, while the 6.75 \m\ emission is directly correlated with the far-infrared emission; the dotted line indicates the (extinction-corrected) inverse one-to-one trend.  Thus, at first glance it would appear that these data suggest there are photospheric contributions to the 4.5 \m\ emission in normal galaxies, at least for those with FIR/B $\lesssim$ 1.  The small number of data points, though, limit the robustness of any such claim.  Finally, if we assume that for higher FIR/B ratios ($\gtrsim5$) the trend levels off, as the results of Helou et al. (2000) and Lu et al. (2000) suggest, then the \ISOcolorc\ ratio for the ``pure ISM'' is about 0.15.

\section{Conclusion}
We present mid-infrared maps at 6.75 and 15 \m\ for 61 normal star-forming galaxies; for a subset of 13 of these galaxies we also show maps at 4.5 \m.  All galaxies for which observations were attempted at these wavelengths were successfully detected.  Qualitatively, the mid-infrared morphology is not a strong function of wavelength, and many of the optical features of the galaxies are also observed in the mid-infrared, except for the bulges of spiral galaxies, consistent with the findings of Helou et al. (1996) in NGC 6946.  Moreover, the data support non-negligible photospheric contributions at 4.5 \m\ for galaxies exhibiting low FIR/B ratios, consistent with the conclusion drawn from mid-infrared ISOPHOT spectroscopy for normal galaxies (Helou et al. 2000; Lu et al. 2000).  However, the evidence presented here is tenuous, as we only have broad band data at 4.5 \m\ for a small number of galaxies.  

Mid-infrared curve of growth profiles indicate that the mid-infrared emission is very well resolved by the ISOCAM maps for most of these galaxies.  Moreover, the profiles are generally exponential in nature, and the distribution of 6.75 and 15 \m\ emission is quite similar for most galaxies.  However, four of the six galaxies that show an enhanced \ISOcolorb\ color also show signs of active central star formation, with yet greater enhancement of \ISOcolorb\ in the central regions.

Quiescent galaxies, those showing low global interstellar heating intensities (i.e. \IRAScolor\ $\geq 0.6$), have an almost constant \ISOcolor\ color near unity.  For galaxies with higher global heating intensities, the mid-infrared color drops rapidly with increasing far-infrared color.  We interpret this as evidence for global mid-infrared spectral energy distributions that are increasingly dominated by \HII\ region emission, characterized by a relatively steep slope in the mid-infrared continuum and a depressed contribution from PAHs.  It is interesting to note that galaxies of all morphological types appear to follow the same color-color trend.

We have estimated the average mid-infrared surface brightness at which the mid-infrared semi-major axis matches that in the optical (at the $B$-band 25 mag arcsec$^{-2}$ level).  We find \RB\ $\approx$ \Rmira\ at 0.04 MJy sr$^{-1}$ and \RB\ $\approx$ \Rmirb\ at 0.09 MJy sr$^{-1}$ on average for this sample.  These mid-infrared surface brightness levels are consistent with observations of Galactic cirrus and the Solar Neighborhood, implying a reasonable similarity in interstellar heating intensity for the outskirts of normal galaxies and our Galaxy.  A final interesting finding from the mid-infrared size analysis centers on the ratio of total heating output to the total heating input for quiescent regions.  We find that ratio to be of order 0.7.

\acknowledgements

This work was supported by \ISO\ data analysis funding from the US National 
Aeronautics and Space Administration, and carried out at the Infrared Processing and Analysis Center (IPAC) and the Jet Propulsion Laboratory of the California Institute of Technology.  The ISOCAM data presented in this paper was analysed using CIA, a joint development by the ESA Astrophysics Division and the ISOCAM
Consortium.  The Digitized Sky Surveys were produced at the Space Telescope Science Institute under U.S. Government grant NAG W-2166. The images of these surveys are based on photographic data obtained using the Oschin Schmidt Telescope on Palomar Mountain and the UK Schmidt Telescope. The plates were processed into the present compressed digital form with the permission of these institutions.

\begin{deluxetable}{lclrlrrr}
\small
\def\pr{$^\prime$}
\tablenum{1}
\tablewidth{42pc}
\tablecaption{U.S. Key Project ISOCAM Galaxies}
\tablehead{
\colhead{}&
\colhead{R.A.(J2000) Dec}&
\colhead{a\tablenotemark{1}} &
\colhead{b\tablenotemark{1}} &
\colhead{}&
\colhead{V$_{\sun}$}&
\colhead{Dist\tablenotemark{2}}&
\colhead{$\log_{10}{{\rm FIR} \over {\rm B}}$\tablenotemark{3}}\\
\colhead{Galaxy}&
\colhead{h:m:s ~~~~~ d:m:s}&
\colhead{$\arcmin$}&
\colhead{$\arcmin$}&
\colhead{Morphology}&
\colhead{km/s}&
\colhead{Mpc}&
\colhead{}
}
\startdata
IC   10          &00:20:24.5 $+$59:17:30&6.3&5.1&IBm?		     & -348 &  0.7 &$-$0.48\nl
NGC  278         &00:52:04.6 $+$47:33:04&2.1&2.0&SAB(rs)b	     &  640 & 11.8 &   0.01\nl
NGC  520         &01:24:35.3   +03:47:37&1.9&0.7&Irr		     & 2266 & 31.6 &   0.36\nl
NGC  693         &01:50:31.0   +06:08:42&2.1&1.0&I0: sp 	     & 1567 & 22.1 &   0.13\nl
NGC  695         &01:51:14.2   +22:34:57&0.8&0.7&IB?(s)m: pec	     & 9735 &131.8 &   0.50\nl
UGC 1449         &01:58:06.7   +03:05:15&1.2&0.7&SBm pec:	     & 5589 & 75.5 &   0.34\nl
NGC  814         &02:10:37.7 $-$15:46:24&1.3&0.5&SB0\^{}0\^{} pec:   & 1616 & 21.4 &   0.54\nl
NGC  986         &02:33:34.1 $-$39:02:41&3.9&3.0&(R\pr)SB(rs)b       & 2005 & 25.2 &   0.25\nl
NGC 1022         &02:38:33.0 $-$06:40:29&2.4&2.0&(R\pr)SB(s)a	     & 1453 & 19.4 &   0.26\nl
UGC 2238         &02:46:17.5   +13:05:44&1.4&1.3&Pec		     & 6436 & 86.8 &   0.82\nl
NGC 1155         &02:58:12.9 $-$10:21:00&0.8&0.7&(R\pr)SAB(s)0o: pec & 4549 & 60.3 &   0.29\nl
NGC 1156         &02:59:42.6   +25:14:17&3.3&2.5&IB(s)m 	     &  375 &  6.4 &$-$0.55\nl
NGC 1222         &03:08:56.8 $-$02:57:18&1.1&0.9&S0-pec:	     & 2452 & 32.6 &   0.44\nl
UGC 2519         &03:09:19.9   +80:07:52&1.2&0.7&SAB?(s:)cd III:     & 2377 & 34.6 &$-$0.04\nl
NGC 1266         &03:16:00.8 $-$02:25:38&1.5&1.0&(R\pr)SB(rs)0 pec   & 2194 & 29.1 &   0.74\nl
NGC 1313         &03:18:15.4 $-$66:29:51&9.1&6.9&SB(s)d 	     &  475 &  3.7 &$-$0.43\nl
NGC 1326         &03:23:56.4 $-$36:27:52&3.9&2.9&(R)SB(rl)0/a	     & 1360 & 16.2 &$-$0.28\nl
NGC 1385         &03:37:28.2 $-$24:30:04&3.4&2.0&SB(s)cd	     & 1493 & 18.4 &$-$0.03\nl
UGC 2855         &03:48:22.6   +70:07:57&4.4&2.0&SB(s)cd II-III      & 1202 & 18.7 &   0.31\nl
NGC 1482         &03:54:39.5 $-$20:30:07&2.5&1.4&SA0+ pec sp	     & 1916 & 24.0 &   0.93\nl
NGC 1546         &04:14:37.2 $-$56:03:35&3.0&1.7&SA?a pec	     & 1276 & 14.1 &$-$0.22\nl
NGC 1569         &04:30:49.0   +64:50:53&3.6&1.8&IBm		     &$-$104&  2.5 &$-$0.66\nl
NGC 2388         &07:28:53.5   +33:49:05&1.0&0.6&SA(s)b: pec	     & 4134 & 54.8 &   1.11\nl
NGC 2366         &07:28:53.7   +69:12:54&8.1&3.3&IB(s)m 	     &  100 &  2.9 &$-$0.83\nl
ESO 317- G 023   &10:24:42.5 $-$39:18:21&1.9&0.8&(R\pr)SB(rs)a       & 2892 & 34.7 &   0.53\nl
IRAS F10565+2448 &10:59:18.1   +24:32:34&0.4&0.3&Pec		     &12921 &171.4 &   1.36\nl
NGC 3583         &11:14:10.8   +48:19:03&2.8&1.8&SB(s)b 	     & 2136 & 29.2 &$-$0.18\nl
NGC 3620         &11:16:04.3 $-$76:12:54&2.8&1.1&(R\pr)SB(s)ab       & 1680 & 19.0 & ....\nl
NGC 3683         &11:27:32.0   +56:52:43&1.9&0.7&SB(s)c?	     & 1716 & 24.2 &   0.42\nl
NGC 3705         &11:30:06.7   +09:16:36&4.9&2.0&SAB(r)ab	     & 1018 & 11.9 &$-$0.62\nl
NGC 3885         &11:46:46.5 $-$27:55:22&2.4&1.0&SAB(r:)0/a:	     & 1802 & 20.8 &   0.04\nl
NGC 3949         &11:53:41.4   +47:51:32&2.9&1.7&SA(s)bc:	     &  807 & 11.6 &$-$0.19\nl
NGC 4102         &12:06:23.1   +52:42:39&3.0&1.7&SAB(s)b?	     &  837 & 12.4 &   0.54\nl
NGC 4194         &12:14:09.6   +54:31:35&1.8&1.1&IBm pec	     & 2506 & 34.8 &   0.63\nl
NGC 4418         &12:26:54.6 $-$00:52:40&1.4&0.7&(R\pr)SAB(s)a       & 2179 & 27.3 &   1.15\nl
NGC 4490         &12:30:36.9   +41:38:23&6.3&3.1&SB(s)d pec	     &  565 &  8.2 &$-$0.20\nl
NGC 4519         &12:33:30.3   +08:39:17&3.2&2.5&SB(rs)d	     & 1220 & 15.1 &$-$0.30\nl
NGC 4713         &12:49:57.8   +05:18:39&2.7&1.7&SAB(rs)d	     &  652 &  7.4 &$-$0.30\nl
IC 3908          &12:56:40.4 $-$07:33:40&1.8&0.7&SB(s)d?	     & 1296 & 15.4 &   0.29\nl
IC  860          &13:15:03.5   +24:37:08&0.5&0.3&SB(s)a:	     & 3347 & 44.7 &   1.04\nl
IC  883          &13:20:35.3   +34:08:22&1.5&1.1&Pec		     & 7000 & 94.0 &   1.14\nl
NGC 5433         &14:02:36.0   +32:30:38&1.6&0.4&SAB(s)c:	     & 4354 & 59.0 &   0.33\nl
NGC 5713         &14:40:11.3 $-$00:17:27&2.8&2.5&SAB(rs)bc pec       & 1883 & 24.7 &   0.20\nl
NGC 5786         &14:58:56.7 $-$42:00:45&2.3&1.1&(R\pr)SAB(s)bc      & 2998 & 37.9 &$-$0.49\nl
NGC 5866         &15:06:29.4   +55:45:49&4.7&1.9&S0		     &  672 & 11.4 &$-$0.61\nl
CGCG 1510.8+0725 &15:13:13.3   +07:13:27&0.4&0.2&SB?(s?)0/a pec      & 3897 & 52.4 &   1.43\nl
NGC 5962         &15:36:31.7   +16:36:32&3.0&2.1&SA(r)c 	     & 1958 & 27.3 &$-$0.09\nl
IC 4595          &16:20:44.2 $-$70:08:35&2.7&0.5&SB?c sp II:	     & 3410 & 42.9 &   0.14\nl
NGC 6286         &16:58:31.4   +58:56:13&1.3&1.2&SB(s)0+ pec?	     & 5501 & 76.5 &   0.83\nl
IC 4662          &17:47:06.4 $-$64:38:25&2.8&1.6&IBm		     &  308 &  2.1 &$-$0.38\nl
NGC 6753         &19:11:23.3 $-$57:02:56&2.5&2.1&(R)SA(r)b	     & 3124 & 40.4 &$-$0.02\nl
NGC 6821         &19:44:24.1 $-$06:50:02&1.2&1.0&SB(s)d:	     & 1525 & 22.5 &$-$0.17\nl
NGC 6822         &19:44:56.1 $-$14:48:05&15.5&13.5&IB(s)m	     &$-$57 &  0.7 &$-$0.58\nl
NGC 6946         &20:34:52.3   +60:09:14&11.5&9.8 &SAB(rs)cd	     &   48 &  5.5 &$-$0.34\nl
NGC 6958         &20:48:42.2 $-$37:59:42&2.1&1.7&E+		     & 2713 & 36.4 &$-$0.88\nl
NGC 7218         &22:10:11.7 $-$16:39:36&2.5&1.1&SB(r)c 	     & 1662 & 23.8 &$-$0.17\nl
NGC 7418         &22:56:36.0 $-$37:01:47&3.5&2.6&SAB(rs)cd	     & 1446 & 19.4 &$-$0.33\nl
IC 5325          &23:28:43.1 $-$41:19:58&2.8&2.5&SAB(rs)bc	     & 1503 & 19.7 &$-$0.26\nl
IRAS F23365+3604 &23:39:01.3   +36:21:10&0.5&0.3&S?Ba? pec or Pec    &19331 &261.2 &   1.16\nl
NGC 7771         &23:51:24.8   +20:06:42&2.5&1.0&SB(s)a 	     & 4277 & 60.0 &   0.41\nl
Markarian  331   &23:51:26.8   +20:35:10&0.7&0.4&SA(s)a: pec	     & 5541 & 76.8 &   1.12\nl
\enddata									     
\tablenotetext{1}{The RC3 major and minor diameters at the $B$-band 25 mag arcsec$^{-2}$ level (de Vaucouleurs et al. 1991).}
\tablenotetext{2}{Distances computed in the Local Group reference frame assuming a Hubble constant of 75 km s$^{-1}$ Mpc$^{-1}$, except for IC 10 and NGC 1569 (see text).}
\tablenotetext{3}{FIR is defined as in Helou et al. (1988): $1.26 \cdot 10^{-14}(2.58I_\nu(60 \mu{\rm m})+I_\nu(100 \mu{\rm m}))$ W m$^{-2}$ where the IRAS flux densities are in units of Jy (see Table \ref{tab:fluxes}).  The monochromatic B-band flux is computed from $\nu I_\nu(0.44 \mu{\rm m})$ where $I_\nu=4260 \cdot 10^{-0.4m_B}$ Jy and $m_B$ is the $k$-corrected face-on B-band apparent magnitude corrected for Galactic extinction (Schlegel, Finkbeiner \& Davis 1998).  We have adopted the RC3 $k$ and internal extinction corrections.}
\label{tab:camsample}								    
\normalsize
\end{deluxetable}								    

\begin{deluxetable}{lrcrr}
\small
\tablenum{2}
\tablewidth{30pc}
\tablecaption{ISOCAM Observations- 2$\times$2 Raster Maps}
\tablehead{
\colhead{}&
\colhead{}&
\colhead{}&
\colhead{}&
\colhead{\hskip-0.7in Integration (sec)}\\
\colhead{Galaxy}&
\colhead{Date}&
\colhead{Rev\tablenotemark{1}}&
\colhead{15 $\mu$m}&
\colhead{6.75 $\mu$m}
}
\startdata
NGC 278 	 & 05 Jul 1997 & 597 & 157 & 151\nl
NGC 520 	 & 01 Jan 1998 & 777\tablenotemark{2} & 157 & 151\nl
NGC 693 	 & 03 Jul 1997 & 595 & 157 & 151\nl
NGC 695 	 & 10 Aug 1997 & 633 & 157 & 151\nl
UGC 1449	 & 24 Jun 1997 & 586 & 157 & 151\nl
NGC 814 	 & 06 Jul 1997 & 598 & 157 & 151\nl
NGC 986 	 & 27 Nov 1997 & 743 & 157 & 151\nl
NGC 1022         & 08 Jan 1998 & 784 & 157 & 151\nl
UGC 2238         & 10 Aug 1997 & 633 & 157 & 151\nl
NGC 1155         & 13 Jul 1997 & 605 & 157 & 151\nl
NGC 1156         & 29 Aug 1997 & 653 & 157 & 151\nl
NGC 1222\tablenotemark{3} & 16 Feb 1998 & 824 &.....& 151\nl
UGC 2519         & 16 Oct 1997 & 701 & 157 & 151\nl
NGC 1266         & 12 Jan 1998 & 788 & 157 & 151\nl	  
NGC 1326         & 04 Dec 1997 & 750\tablenotemark{2} & 157 & 151\nl   
NGC 1385         & 19 Jan 1998 & 796 & 157 & 151\nl
UGC 2855         & 06 Aug 1997 & 629 & 157 & 151\nl
NGC 1482         & 19 Jan 1998 & 796\tablenotemark{2} & 157 & 151\nl
NGC 1546         & 04 Oct 1997 & 689 & 157 & 151\nl
NGC 1569         & 27 Mar 1998 & 863\tablenotemark{4} & 157 & 151\nl
NGC 2388         & 03 Nov 1997 & 718\tablenotemark{2} & 157 & 151\nl
ESO 317- G 023   & 25 Jul 1996 & 252 & 63 & 121 \nl
IRAS F10565+2448 & 05 Jun 1996 & 202 & 63 & 121 \nl
NGC 3583         & 30 May 1996 & 195 & 63 & 121 \nl
NGC 3620         & 18 Aug 1996 & 276 & 63 & 121 \nl
NGC 3620         & 21 Sep 1997 & 675\tablenotemark{2}&.....&.....\nl
NGC 3683         & 29 May 1996 & 194 & 63 & 121\nl
NGC 3705         & 19 May 1996 & 184 & 63 & 121\nl
NGC 3885         & 26 Jul 1996 & 252 & 63 & 121\nl
NGC 3949         & 30 May 1996 & 195 & 63 & 121\nl
NGC 4102         & 30 May 1996 & 195 & 63 & 121\nl
NGC 4194         & 29 May 1996 & 194 & 63 & 121\nl
NGC 4418         & 14 Jul 1996 & 241 & 63 & 121\nl
NGC 4490         & 09 Jun 1996 & 205 & 63 & 121\nl
NGC 4519         & 10 Jul 1996 & 236 & 63 & 121\nl
NGC 4713         & 29 Jun 1996 & 225 & 63 & 121\nl
IC 3908          & 26 Jul 1996 & 252 & 63 & 121\nl
IC 860           & 16 Jul 1996 & 243 & 63 & 121\nl
IC 883           & 19 Jun 1996 & 215 & 63 & 121\nl
NGC 5433         & 09 Jun 1997 & 571 &157 & 151\nl
NGC 5713         & 26 Aug 1996 & 284 & 63 & 121\nl
NGC 5713         & 05 Aug 1997 & 628\tablenotemark{2}&.....&.....\nl
NGC 5786         & 10 Sep 1996 & 299 & 63 & 121\nl
NGC 5866         & 08 Aug 1997 & 631\tablenotemark{2}&.....&.....\nl
CGCG 1510.8+0725 & 24 Aug 1996 & 281 & 63 & 121\nl
NGC 5962         & 20 Aug 1996 & 278 & 63 & 121\nl
IC 4595          & 19 Aug 1996 & 276 & 63 & 121\nl
IC 4595          & 03 Aug 1997 & 626\tablenotemark{2}&.....&.....\nl
NGC 6286         & 11 Jun 1996 & 207 & 63 & 121\nl
IC 4662          & 25 Aug 1997 & 648 & 63 & 121\nl
NGC 6753         & 10 Sep 1996 & 299 & 63 & 121\nl
NGC 6753         & 31 Oct 1997 & 716\tablenotemark{2}&.....&.....\nl
NGC 6821         & 30 Sep 1996 & 319 & 63 & 121\nl
NGC 6958         & 11 May 1997 & 542 &157 & 151\nl
NGC 7218         & 19 Nov 1996 & 369 & 63 & 121\nl
NGC 7418         & 19 Nov 1996 & 369 & 63 & 121\nl
IC 5325          & 20 Nov 1996 & 369 & 63 & 121\nl
IRAS F23365+3604 & 07 Dec 1997 & 752 &157 & 151\nl
NGC 7771         & 03 Jun 1997 & 565 & 63 & 121\nl
NGC 7771         & 01 Dec 1997 & 747\tablenotemark{2}&.....&.....\nl
Markarian 331    & 03 Jun 1997 & 565 & 63 & 121\nl
\enddata
\tablenotetext{1}{Revolution number (orbit) in which the data were obtained}
\tablenotetext{2}{A 4.5 \m\ map was obtained during this revolution.
The integration time was 302 seconds for a 4.5 \m\ map.}
\tablenotetext{3}{The 15 \m\ observation for NGC 1222 failed.}
\tablenotetext{4}{NGC 1569 was observed using several of the ISOCAM filters.
More details are given in the text and in Table \ref{tab:n1569}.}
\label{tab:small_maps}
\normalsize
\end{deluxetable}

\begin{deluxetable}{lcrrrcccc}
\small
\def\tim{$\times$}
\tablenum{3}
\tablewidth{35pc}
\tablecaption{ISOCAM Observations-Large Maps}
\tablehead{
\colhead{}&
\colhead{}&
\colhead{}&
\multicolumn{3}{c}{Integration (seconds)}\\
\colhead{Galaxy}&
\colhead{Date}&
\colhead{Rev\tablenotemark{1}}&
\colhead{15 $\mu$m}&
\colhead{6.75 $\mu$m}&
\colhead{4.5 $\mu$m}&
\colhead{rastering}&
\colhead{Size}
}
\startdata
IC 10     & 15 Feb 1997 & 457 &  504 &  605 & .....& 4\tim4 & ~7\farcm25\tim~7\farcm25\nl
IC 10\tablenotemark{2}& 16 Feb 1998 & 841 & 630 & 605 & 1210 & 4\tim4 & ~7\farcm25\tim~7\farcm25\nl
NGC 1313  & 18 Apr 1997 & 519 & 1008 & 1210 & .....&2\tim4\tim4 & 11\farcm85\tim~9\farcm20 \nl
NGC 1313  & 08 Sep 1997 & 663 & .....& .....& 2419 &2\tim4\tim4 & ~9\farcm20\tim11\farcm85 \nl
NGC 2366  & 04 Sep 1997 & 658 &  840 & 1008 & .....& 5\tim5 & ~8\farcm20\tim~8\farcm20 \nl
NGC 6822  & 03 Nov 1997 & 719 & 2352 & 2822 & .....& 8\tim8 & 12\farcm65\tim12\farcm65 \nl
NGC 6946  & 08 Feb 1996 &  83 & 1176 & 2258 & .....& 8\tim8 & 12\farcm65\tim12\farcm65\nl
\enddata
\tablenotetext{1}{Revolution number (orbit) in which the data were obtained.}
\tablenotetext{2}{IC 10 was also observed at 11.4 \m\ during this revolution for 605 seconds.}
\label{tab:large_maps}
\normalsize
\end{deluxetable}

\begin{deluxetable}{lrrrrrr}
\small
\def\t2{\tablenotemark{2}}
\def\p{$\pm$}
\tablenum{4}
\tablewidth{35pc}
\tablecaption{Flux Densities}
\tablehead{
\colhead{}&\colhead{12 \m}&\colhead{25 \m}&
\colhead{60 \m}&\colhead{100 \m}&
\colhead{6.75 \m}&\colhead{15 \m}\\
\colhead{Galaxy}&\colhead{Jy}&\colhead{Jy}&\colhead{Jy}&\colhead{Jy}&
\colhead{Jy}&\colhead{Jy}
}
\startdata
IC  10   &  4.88 & 13.95 & 112.92 &178.56 &   2.18\p0.44~ &   3.06\p0.61~\nl
NGC  278 &  1.63 &  2.57 &  25.05 & 46.22 &   1.12\p0.23~ &   1.18\p0.24~\nl
NGC  520 &  0.76 &  2.84 &  31.10 & 46.99 &   0.52\p0.13: &   0.86\p0.18~\nl
NGC  693 &  0.29 &  0.55 &   6.73 & 11.83 &   0.18\p0.04~ &   0.18\p0.05~\nl
NGC  695 &  0.48 &  0.86 &   7.87 & 13.62 &   0.25\p0.05~ &   0.36\p0.09~\nl
UGC 1449 &  0.31 &  0.56 &   4.96 &  8.39 &   0.18\p0.04~ &   0.17\p0.03~\nl
NGC  814 &  0.19 &  1.01 &   4.41 &  3.60 &   0.05\p0.01~ &   0.22\p0.05~\nl
NGC  986 &  1.41 &  3.65 &  25.14 & 51.11 &   0.51\p0.13: &   1.07\p0.25~\nl
NGC 1022 &  0.75 &  3.29 &  19.83 & 27.09 &   0.31\p0.08: &   0.77\p0.16~\nl
UGC 2238 &  0.34 &  0.53 &   8.40 &  0.54 &   0.24\p0.05~ &   0.30\p0.07~\nl
NGC 1155 &  0.21 &  0.47 &   2.89 &  5.01 &   0.05\p0.01~ &   0.09\p0.02~\nl
NGC 1156 &  0.17 &  0.55 &   5.24 & 10.48 &   0.09\p0.02~ &   0.14\p0.03~\nl
NGC 1222 &  0.51 &  2.29 &  13.07 & 15.45 &   0.27\p0.06~ &   ..........~\nl
UGC 2519 &  0.25 &  0.34 &   2.98 &  7.46 &   0.12\p0.03~ &   0.17\p0.04~\nl
NGC 1266 &  0.14 &  1.23 &  13.32 & 16.43 &   0.04\p0.01~ &   0.15\p0.03~\nl
NGC 1313 &  1.70 &  3.75 &  45.69 & 97.25 &   0.69\p0.14~ &   0.84\p0.17~\nl
NGC 1326 &  0.38 &  0.86 &   8.17 & 13.90 &   0.18\p0.04~ &   0.32\p0.07~\nl
NGC 1385 &  1.19 &  2.02 &  17.30 & 37.23 &   0.72\p0.15~ &   0.76\p0.16~\nl
UGC 2855\t2&2.93 &  4.86 &  42.39 & 90.49 &$>$2.15\p0.54~~&$>$2.39\p0.60~\nl
NGC 1482 &  1.54 &  4.67 &  33.45 & 46.53 &   0.78\p0.20: &   1.37\p0.28~\nl
NGC 1546 &  0.62 &  0.79 &   7.21 & 22.44 &   0.42\p0.08~ &   0.56\p0.12~\nl
NGC 1569 &  1.23 &  8.98 &  54.25 & 55.30 &   0.32\p0.06~ &   1.43\p0.29~\nl
NGC 2388 &  0.51 &  2.07 &  17.01 & 25.33 &   0.23\p0.05~ &   0.50\p0.11~\nl
NGC 2366 &  0.21 &  1.05 &   4.85 &  5.04 &   0.02\p0.005~&   0.17\p0.04~\nl
ESO 317-G 023&0.34& 0.88 &  13.50 & 23.71 &   0.18\p0.04~ &   0.27\p0.06~\nl
IRAS F10565+2448&0.17&1.19& 12.08 & 15.36 &   0.09\p0.02~ &   0.27\p0.08~\nl
NGC 3583 &  0.63 &  0.78 &   7.08 & 18.66 &   0.29\p0.06~ &   0.41\p0.08~\nl
NGC 3620 &  1.29 &  4.71 &  46.80 & 67.26 &   0.63\p0.16: &   1.06\p0.22~\nl
NGC 3683 &  1.06 &  1.53 &  13.61 & 29.27 &   0.64\p0.13~ &   0.76\p0.16~\nl
NGC 3705 &  0.38 &  0.44 &   3.72 & 11.22 &   0.19\p0.05~ &   0.27\p0.09~\nl
NGC 3885 &  0.46 &  1.41 &  11.66 & 16.46 &   0.33\p0.08: &   0.45\p0.10~\nl
NGC 3949 &  0.82 &  1.37 &  11.28 & 26.65 &   0.42\p0.09~ &   0.50\p0.12~\nl
NGC 4102 &  1.72 &  7.05 &  48.10 & 71.01 &   0.66\p0.16: &   1.61\p0.40:\nl
NGC 4194 &  0.83 &  4.53 &  23.72 & 25.44 &   0.41\p0.10: &   0.84\p0.17~\nl
NGC 4418 &  1.00 &  9.69 &  43.89 & 31.94 &   0.24\p0.06: &   1.56\p0.39:\nl
NGC 4490\t2&1.86 &  4.20 &  45.90 & 76.93 &$>$1.07\p0.27~~&$>$1.40\p0.35~\nl
NGC 4519 &  0.36 &  0.55 &   3.74 &  7.02 &   0.11\p0.02~ &   0.23\p0.05~\nl
NGC 4713 &  0.24 &  0.17 &   4.60 & 10.84 &   0.15\p0.03~ &   0.21\p0.05~\nl
IC  3908 &  0.44 &  0.87 &   8.09 & 17.08 &   0.44\p0.10~ &   0.43\p0.12~\nl
IC   860 &  0.10 &  1.31 &  17.93 & 18.60 &   0.02\p0.004~&   0.06\p0.02~\nl
IC   883 &  0.25 &  1.41 &  17.01 & 24.41 &   0.16\p0.03~ &   0.25\p0.05~\nl
NGC 5433 &  0.27 &  0.70 &   6.62 & 11.57 &   0.18\p0.04~ &   0.30\p0.07~\nl
NGC 5713 &  1.30 &  2.84 &  21.89 & 38.09 &   0.88\p0.18~ &   1.06\p0.22~\nl
NGC 5786 &  0.36 &  0.76 &   5.26 & 14.98 &   0.30\p0.07~ &   0.39\p0.11~\nl
NGC 5866 &  0.36 &  0.34 &   5.21 & 17.11 &   ..........~ &   ..........~\nl
CGCG1510.8+0725&0.05&0.83&  20.84 & 31.52 &   0.04\p0.01~ &   0.08\p0.03~\nl
NGC 5962 &  0.74 &  1.03 &   8.89 & 22.11 &   0.37\p0.08~ &   0.51\p0.12~\nl
IC  4595 &  0.71 &  0.73 &   7.05 & 18.04 &   0.30\p0.07~ &   0.39\p0.12~\nl
NGC 6286 &  0.42 &  0.56 &   8.22 & 22.13 &   0.17\p0.03~ &   0.20\p0.04~\nl
IC  4662 &  0.30 &  1.27 &   8.81 & 11.90 &   0.05\p0.02~ &   0.17\p0.04~\nl
NGC 6753 &  0.95 &  0.98 &   9.77 & 28.26 &   0.38\p0.08~ &   0.51\p0.12~\nl
NGC 6821 &  0.14 &  0.31 &   3.63 &  5.71 &   0.06\p0.01~ &   0.13\p0.04~\nl
NGC 6822 &  0.84 &  6.63 &  58.86 &130.32 &   0.10\p0.02~ &   0.31\p0.06~\nl
NGC 6946 & 15.17 & 23.34 & 167.72 &362.66 &  11.19\p2.24~ &  11.49\p2.30~\nl
NGC 6958 &  0.16 &  0.20 &   1.00 &  1.99 &   0.03\p0.01~ &   0.03\p0.01~\nl
NGC 7218 &  0.28 &  0.56 &   4.67 & 11.18 &   0.21\p0.04~ &   0.28\p0.07~\nl
NGC 7418 &  0.63 &  0.69 &   5.38 & 16.13 &   0.23\p0.05~ &   0.29\p0.07~\nl
IC  5325 &  0.48 &  0.70 &   5.15 & 14.35 &   0.24\p0.06~ &   0.31\p0.08~\nl
IRAS 23365+3604&0.13&0.88&   7.44 &  8.83 &   0.02\p0.005~&   0.11\p0.02~\nl
NGC 7771 &  0.77 &  1.77 &  19.67 & 40.12 &   0.46\p0.11: &   0.59\p0.12~\nl
Mark 0331&  0.55 &  2.39 &  18.04 & 23.61 &   0.21\p0.05: &   0.52\p0.11~\nl
\enddata
\tablenotetext{1}{Fluxes with a colon indicate ADC saturation problems.  These flux densities are lower limits.}
\tablenotetext{2}{We estimate the total uncertainty in the flux density measures for NGC 4490 and UGC 2855 to be somewhat larger than the sample average uncertainty, since these galaxies are large compared to our mid-infrared maps.}
\label{tab:fluxes}
\normalsize
\end{deluxetable}

\newpage
\begin{deluxetable}{lr}
\small
\def\p{$\pm$}
\tablenum{5}
\tablewidth{12pc}
\tablecaption{4.5 \m\ Fluxes}
\tablehead{
\colhead{Galaxy}&
\colhead{Flux}\\
\colhead{}&\colhead{Jy}
}
\startdata
IC    10& 0.45\p0.11\nl 
NGC  520& 0.07\p0.02\nl 
NGC 1313& 0.10\p0.03\nl
NGC 1326& 0.09\p0.02\nl
NGC 1482& 0.10\p0.02\nl
NGC 1569& 0.11\p0.03\nl
NGC 2388& 0.03\p0.01\nl
NGC 3620& 0.14\p0.03\nl 
NGC 5713& 0.10\p0.03\nl 
NGC 5866& 0.22\p0.05\nl 
IC  4595& 0.05\p0.02\nl 
NGC 6753& 0.14\p0.03\nl 
NGC 7771& 0.12\p0.03\nl 
\enddata
\label{tab:4.5mu}
\normalsize
\end{deluxetable}

\begin{deluxetable}{crr}
\small
\def\p{$\pm$}
\tablenum{6}
\tablewidth{12pc}
\tablecaption{NGC 1569 Fluxes}
\tablehead{
\colhead{Filter}&
\colhead{$\lambda$}&
\colhead{Flux}\\
\colhead{}&\colhead{\m}&\colhead{Jy}
}
\startdata
LW1&  4.5& 0.11\p0.03 \nl
LW4&  6.0& 0.15\p0.03 \nl
LW2&  6.7& 0.32\p0.06 \nl
LW6&  7.7& 0.40\p0.08 \nl
LW7&  9.6& 0.44\p0.09 \nl
LW8& 11.3& 0.78\p0.16 \nl
LW3& 15.0& 1.43\p0.29 \nl
LW9& 14.9& 1.82\p0.36 \nl
\enddata
\label{tab:n1569}
\normalsize
\end{deluxetable}

\begin{figure}
\caption{Optical (from the Digitized Sky Survey) and mid-infrared maps for the ISOCAM sample.  The first frame is from the DSS.  The second frame, if available, is the 4.5 \m\ map, and the third and fourth frames are the 6.75 and 15 \m\ maps.  The coordinate system included in the DSS frame indicates the N-E orientation for the DSS, 6.75 and 15 \m\ frames, with the arrow pointing towards N.  Some of the 4.5 \m\ observations were done at a different roll angle, so we have also included the N-E orientation for those observations.}
\label{fig:cam_images}
\end{figure}

\begin{figure}
\caption{Digitized Sky Survey optical and ISOCAM mid-infrared maps at 4.5, 6.75, 11.4, and 15 \m\ of IC 10.  The coordinate system included in the DSS frame indicates the N-E orientation, with the arrow pointing towards N.}
\label{fig:ic10}
\end{figure}

\begin{figure}
\caption{Digitized Sky Survey optical and ISOCAM mid-infrared maps at 4.5, 6.75, and 15 \m\ of NGC 1313.  The coordinate system included in the DSS frame indicates the N-E orientation, with the arrow pointing towards N.  The roll angle for the 4.5 \m\ observations were done at a different roll angle, so we have rotated the image for N-E orientation consistency.}
\label{fig:n1313}
\end{figure}

\begin{figure}
\caption{Digitized Sky Survey optical and ISOCAM mid-infrared maps at 6.75 and 15 \m\ of NGC 2366.  The coordinate system included in the DSS frame indicates the N-E orientation, with the arrow pointing towards N.}
\label{fig:n2366}
\end{figure}

\begin{figure}
\caption{Digitized Sky Survey optical and ISOCAM mid-infrared maps at 6.75 and 15 \m\ of NGC 6822.  The coordinate system included in the DSS frame indicates the N-E orientation, with the arrow pointing towards N.}
\label{fig:n6822}
\end{figure}

\begin{figure}
\caption{Digitized Sky Survey optical and ISOCAM mid-infrared maps at 6.75 and 15 \m\ of NGC 6946.  The coordinate system included in the DSS frame indicates the N-E orientation, with the arrow pointing towards N.}
\label{fig:n6946}
\end{figure}

\end{document}